\documentclass[journal]{IEEEtran}
\IEEEoverridecommandlockouts
\usepackage{cite}
\usepackage{amsmath,amssymb,amsfonts}
\usepackage{algorithmic}
\usepackage{graphicx}
\usepackage{textcomp}
\usepackage{xcolor}
\usepackage{xpatch}
\usepackage{amsfonts}
\usepackage{bm}
\usepackage{booktabs}
\usepackage{algorithm}
\usepackage{multirow,bm,bbm,array,setspace}
\usepackage{textcomp}
\usepackage{psfrag}
\usepackage{pstricks,enumerate}
\usepackage{bbm}

\makeatletter
\newcommand{\distas}[1]{\mathbin{\overset{#1}{\kern\z@\sim}}}%
\newsavebox{\mybox}\newsavebox{\mysim}
\newcommand{\distras}[1]{%
  \savebox{\mybox}{\hbox{\kern1pt$\scriptstyle#1$\kern1pt}}%
  \savebox{\mysim}{\hbox{$\sim$}}%
  \mathbin{\overset{#1}{\kern\z@\resizebox{\wd\mybox}{\ht\mysim}{$\sim$}}}%
}
\ExplSyntaxOn
\cs_new:Npn \bibColoredItems #1#2
  {
    \clist_map_inline:nn {#2} { \cs_new:cpn {bib@colored@##1} {#1} } 
  }
\ExplSyntaxOff

\newcommand\bib@setcolor[1]{%
  \ifcsname bib@colored@#1\endcsname
    \expanded{\noexpand\color{\csname bib@colored@#1\endcsname}}%
  \else
    \normalcolor
  \fi
}

\IfPackageLoadedTF{hyperref}{\@tempswatrue}{\@tempswafalse}
\if@tempswa
  \xpatchcmd\@bibitem {\H@item}{\bib@setcolor{#1}\H@item}{}{\PatchFailed}
  \xpatchcmd\@lbibitem{\H@item}{\bib@setcolor{#2}\H@item}{}{\PatchFailed}
\else
  \xpatchcmd\@bibitem {\item}  {\bib@setcolor{#1}\item}  {}{\PatchFailed}
  \xpatchcmd\@lbibitem{\item}  {\bib@setcolor{#2}\item}  {}{\PatchFailed}
\fi

\makeatother
\ifCLASSINFOpdf
\else
\fi
\newcommand{\bp}{\bm{p}}
\newcommand{\bS}{\bm{S}}
\newcommand{\bs}{\bm{s}}
\newcommand{\bI}{\bm{I}}

\newcommand{\bV}{\bm V}

\newcommand{\bU}{\bm U}
\newcommand{\bY}{\bm{Y}}
\newcommand{\by}{\bm{y}}
\newcommand{\bz}{\bm{z}}
\newcommand{\bZ}{\bm{Z}}
\newcommand{\bQ}{\bm{Q}}
\newcommand{\bx}{\bm{x}}
\newcommand{\bA}{\bm{A}}
\newcommand{\ba}{\bm{a}}
\newcommand{\bB}{\bm{B}}
\newcommand{\bG}{\bm{G}}
\newcommand{\bM}{\bm{M}}
\newcommand{\bF}{\bm{F}}
\newcommand{\bbf}{\bm{f}}
\newcommand{\bK}{\bm{K}}

\newtheorem{proposition}{Proposition}

\newtheorem{theorem}{Theorem}

\newtheorem{remark}{Remark}

\begin{document}

\title{Mixed Max-and-Min Fractional Programming for Wireless Networks}
\author{
\IEEEauthorblockN{
Yannan Chen, \IEEEmembership{Student Member,~IEEE}, Licheng Zhao, and Kaiming Shen, \IEEEmembership{Member,~IEEE}
} 
\thanks{Manuscript accepted to IEEE Transactions on Signal Processing December 13, 2023.
This work has been presented in part at IEEE International Conference on Acoustics, Speech, and Signal Processing
 (ICASSP), Rhodes Island, Greece, 2023, and in part at IEEE International Workshop on Signal Processing Advances in Wireless Communications (SPAWC), Shanghai, China, 2023. \emph{(Corresponding author: Kaiming Shen.)}
 
Yannan Chen and Kaiming Shen are with School of Science and Engineering, The Chinese University of Hong Kong (Shenzhen), Shenzhen 518172, China (e-mail: yannanchen@link.cuhk.edu.cn; shenkaiming@cuhk.edu.cn).

Licheng Zhao is with Shenzhen Research Institute of Big Data, Shenzhen 518172, China (e-mail: zhaolicheng@sribd.cn).
}
}

\maketitle

\begin{abstract}
Fractional programming (FP) plays a crucial role in wireless network design because many relevant problems involve maximizing or minimizing ratio terms. Notice that the maximization case and the minimization case of FP cannot be converted to each other in general, so they have to be dealt with separately in most of the previous studies. Thus, an existing FP method for maximizing ratios typically does not work for the minimization case, and vice versa. However, the FP objective can be mixed max-and-min, e.g., one may wish to maximize the signal-to-interference-plus-noise ratio (SINR) of the legitimate receiver while minimizing that of the eavesdropper. We aim to fill the gap between max-FP and min-FP by devising a unified optimization framework. The main results are three-fold. First, we extend the existing max-FP technique called quadratic transform to the min-FP, and further develop a full generalization for the mixed case. Second, we provide a minorization-maximization (MM) interpretation of the proposed unified approach, thereby establishing its convergence and also obtaining a matrix extension; another result we obtain is a generalized Lagrangian dual transform which facilitates the solving of the logarithmic FP. Finally, we present three typical applications: the age-of-information (AoI) minimization, the Cram\'{e}r-Rao bound minimization for sensing, and the secure data rate maximization, none of which can be efficiently addressed by the previous FP methods.
\end{abstract}

\begin{IEEEkeywords}
Multi-ratio mixed max-and-min fractional programming (FP); matrix FP; unified approach; age-of-information (AoI); Cram\'{e}r-Rao bound; sensing; secure transmission.
\end{IEEEkeywords}

\section{Introduction}
\IEEEPARstart{I}{n} a broad sense, fractional programming (FP) refers to a class of mathematical optimizations whose main components are ratios. The existing studies in this area mostly focus on either the \emph{concave-convex sum-of-ratios max-FP} problem
\begin{align}
\label{prob:FP_max}
  \underset{\bx\in\mathcal X}{\text{maximize}} &\quad \sum^N_{n=1}\frac{A_n(\bx)}{B_n(\bx)},
\end{align}
where each $A_n(\bx)\ge0$ is a  concave nonnegative function, each $B_n(\bx)>0$ is a  convex positive function, and $\mathcal X$ is a nonempty convex constraint set, or the min-FP version
\begin{align}
\label{prob:FP_min}
\underset{\bx\in\mathcal X}{\text{minimize}} &\quad \sum^N_{n=1}\frac{A_n(\bx)}{B_n(\bx)}
\end{align}
with the same setting as in \eqref{prob:FP_max} except that the convexity and the concavity assumptions are swapped between $A_n(\bx)$ and $B_n(\bx)$. Notice that neither \eqref{prob:FP_max} nor \eqref{prob:FP_min} is a convex problem in general even if the objective function contains only one ratio.

As a classical result on FP, even though the single-ratio case of \eqref{prob:FP_max} with $N=1$ is nonconvex, it can be efficiently solved by the Charnes-Cooper algorithm 
\cite{charnes1962programming,schaible1974parameter} or by Dinkelbach's algorithm \cite{dinkelbach1967nonlinear}. Moreover, since minimizing a ratio is equivalent to maximizing its reciprocal, the single-ratio min-FP in \eqref{prob:FP_max} with $N=1$ can be rewritten in the max-FP form by flipping the ratio; thereafter, the classical FP methods in \cite{charnes1962programming,schaible1974parameter,dinkelbach1967nonlinear} apply immediately. But such reciprocal equivalence between max-FP and min-FP no longer holds in the presence of multiple ratios, so they are still examined separately in most of the existing literature. In contrast, we set out a unified approach.

In the FP field, Dinkelbach's algorithm \cite{dinkelbach1967nonlinear} has long been the most popular approach to the single-ratio problem because of its superlinear convergence to the global optimum. Dinkelbach's algorithm can be extended to 
the maxmin FP problem with multiple ratios \cite{crouzeix1985algorithm}. Nevertheless, efficiently finding the global optimum of \eqref{prob:FP_max} or \eqref{prob:FP_min} remains an open problem.
An early attempt in \cite{almogy1971class} suggested transforming \eqref{prob:FP_max} into the problem of finding roots of a monotone-decreasing convex function, but it has been shown invalid in \cite{falk1992optimizing} even if $A_n(\bx)$'s and $B_n(\bx)$'s are all linear functions. Actually, as established in \cite{gruzdeva2018solving}, the difficulty of solving \eqref{prob:FP_min} lies in the fact that the problem amounts to minimizing the difference between two convex functions---which is notoriously difficult. Furthermore, \cite{freund2001solving} shows that 
\eqref{prob:FP_max} and \eqref{prob:FP_min} are both NP-complete. As a result, quite a few existing works resort to branch and bound (B\&B), e.g., \cite{freund2001solving,thi2003unified,konno2000branch,benson2001global,qu2007efficient} about max-FP and  \cite{kuno2002branch,liu2019new,benson2007solving,benson2002global,benson2002using} about min-FP. Again, since max-FP and min-FP are fundamentally different, the B\&B procedure for max-FP typically does not work for min-FP, and vice versa. To avoid the exponential time complexity, other works are based upon a variety of heuristic algorithms, e.g., the ``state space reduction'' method \cite{falk1992optimizing} and the ``harmony search'' method \cite{jaberipour2010solving}.  {Our work aims to find a provable ``good'' solution (i.e., a stationary point) efficiently.}

Regarding the practical use of FP in wireless network design, energy efficiency maximization is arguably the earliest application considered by the community. A survey on this topic may be found in \cite{zappone2015energy}.
{Aside from the max-min energy efficiency problem \cite{zappone2015energy} being optimally solved by the generalized Dinkelbach's algorithm \cite{crouzeix1985algorithm},
the sequential-FP framework based on Dinkelbach's algorithm \cite{zappone2017globally} and the first-order optimal method \cite{matthiesen2020globally} have been developed to cope with the multi-link energy efficiency problems.} 
% { In contrast to Dinkelbach's transform, \cite{matthiesen2020globally} considers a more general energy efficiency problem and proposes a faster branch and bound procedure to find the global optimal, a successive convex approximation method to find the local optimal with practical complexity, and train a neural network to predict the optimal resource allocation. While \cite{matthiesen2020globally} can only tackle the energy efficiency problem,} 
The more recent technique called quadratic transform \cite{shen2018fractional} aims at a broad range of multi-ratio max-FP problems. 
{The above methods \cite{matthiesen2020globally,zappone2017globally,shen2018fractional} are all based on the monotonic improvement of the objective value and can guarantee convergence to a stationary point of the max-FP problem.}
Equipped with the quadratic transform, FP {has been considered for a variety of application instances of wireless system design}, e.g., intelligent reflecting surface (IRS) or reconfigurable intelligent surface (RIS) \cite{liu2020intelligent}%,mamaghani2022aerial,xu2022intelligent}
, hybrid beamforming \cite{li2019hybrid}, %ni2022user}, 
wireless power transfer %\cite{xu2019robust,ye2021joint,}
\cite{ren2023transmission}, edge computing \cite{ding2021joint,ding2020joint,reifert2023rate}, unmanned aerial vehicle (UAV) \cite{feng2020joint},
%shafique2020optimization,jiang2019joint}, 
 massive multiple-input multiple-output (MIMO) \cite{shen2020enhanced}, %\cite{sharma2019energy,shen2020enhanced,mandawaria2020energy}, 
 integrated sensing and communication (ISAC) \cite{keskin2021limited}, %,dong2020low,ding2022joint},
etc. In principle, the quadratic transform promises an efficient algorithm whenever the practical problem can be formulated as maximizing a sum of positive ratios as in \eqref{prob:FP_max}.

However, we would like to point out that the quadratic transform \cite{shen2018fractional} only works for the max-FP case, and also that the ratios are not always to be maximized in wireless networks. For instance, the age-of-information (AoI), which is a common metric for quantifying the freshness of data packets \cite{kaul2011minimizing}, turns out to be fractionally structured; a widely considered problem is to coordinate the packet-generating rates of multiple sources in order to minimize the total AoI. Thus, we end up with a min-FP problem. A similar problem scenario would be encountered when we aim to minimize the average Cram\'{e}r-Rao bound \cite{van2002optimum} on the sensing error by coordinating waveforms across the radar sets. Furthermore, max-FP and min-FP may even coexist, e.g., when one wishes to maximize the signal-to-interference-plus-noise ratio (SINR) of the legitimate receiver while minimizing that of the eavesdropper \cite{kalantari2015joint}. The above three applications constitute a selection of optimization problems that can be quite difficult if considered without using the new FP technique as proposed in this paper. %We have no intention to provide a complete survey on the possible applications of our new FP method, just mentioning that it applies as well to the latency minimization problem \cite{park2021collaborative}, the exposure dose problem \cite{heliot2022exposure}, and integrated sensing and communications \cite{liu2021cramer}.

Below is a summary of the main contributions of this work:
\begin{itemize}
    \item \emph{New FP Technique:} We extend the existing max-FP technique called quadratic transform \cite{shen2018fractional} to the min-FP case and then to the mixed max-and-min FP case. This unified approach can be further developed to account for matrix ratios. Moreover, we devise a generalized Lagrangian dual transform for the logarithmic FP problem---which can be frequently encountered in wireless network design.
    \item \emph{MM interpretation:} We provide analytical justification of the proposed unified approach by connecting it to the minorization-maximization (MM) theory. In light of the MM interpretation, we show that the iterative optimization by the unified approach yields a monotonic improvement, and further yields convergence to a stationary point under certain conditions.
    \item \emph{Applications:} We discuss three typical applications of the proposed unified approach: (i) the AoI minimization which represents the min-FP case, (ii) the Cram\'{e}r-Rao bound minimization for sensing which represents the matrix min-FP case, and (iii) the secure data rate maximization which represents the mixed max-and-min FP case. None of the above three problems can be efficiently addressed by the previous FP methods.
\end{itemize}

The rest of the paper is organized as follows. Section \ref{sec:FP} reviews the quadratic transform \cite{shen2018fractional} for the max-FP, and then proposes its extensions to the min-FP and the mixed max-and-min FP. Section \ref{sec:MM} examines the connection between the proposed unified approach and the MM theory, based on which we further discuss the matrix extension and the generalized Lagrangian dual transform. Section \ref{sec:app} presents three application cases. Finally, Section \ref{sec:conclusion} concludes the paper.

Here and throughout, bold lower-case letters represent vectors while bold upper-case letters represent matrices. 
% and the calligraphy upper-case letter a set. 
For matrix $\bA$, $\bA^{\top}$ is its transpose, $\bA^{\text{H}}$ is its conjugate transpose, and $\bA^{-1}$ is its inverse. For vector $\ba$, $\ba^{\top}$ is its transpose and $\ba^{\text{H}}$ is its conjugate transpose. The set of real numbers, the set of nonnegative numbers, and the set of strictly positive numbers are denoted by $\mathbb R$, $\mathbb R_+$, and $\mathbb R_{++}$, respectively. The set of complex numbers, the set of Hermitian positive semi-definite matrices, and the set of Hermitian positive definite matrices are denoted by $\mathbb{C}$, 
$\mathbb{H}_{+}$, and $\mathbb{H}_{++}$, respectively. Moreover, $\bA\succ\bB$ indicates that $(\bA-\bB)\in\mathbb H_{++}$, $\mathrm{vec}(\bA)$ is the vectorization of matrix $\bA$, and $\otimes$ is the Kronecker product. {For a real number $a\in\mathbb R$, define $[a]_+ = \lim_{\epsilon\rightarrow0_+}\max\{a,\epsilon\}$}.

\section{Fractional Programming}
\label{sec:FP}

In this section we start by reviewing the existing max-FP technique---the quadratic transform \cite{shen2018fractional}, which can address a broader range of multi-ratio problems than the traditional Charnes-Cooper algorithm \cite{charnes1962programming,schaible1974parameter} and Dinkelbach's algorithm \cite{dinkelbach1967nonlinear}. The first main result of the paper is a new min-FP method called \emph{inverse quadratic transform}. Most importantly, it can be integrated with the above quadratic transform to obtain a unified approach to the mixed max-and-min FP.

\subsection{Max-FP Case}

Following the traditional methods such as Charnes-Cooper algorithm \cite{charnes1962programming,schaible1974parameter} and Dinkelbach's algorithm \cite{dinkelbach1967nonlinear}, the primary idea of the quadratic transform \cite{shen2018fractional} is to separate the numerator and denominator of each ratio term, but its ratio decoupling strategy works for a broader range of max-FP problems including the sum-of-ratios case, as stated below.

%We start with the quadratic transform proposed in \cite{shen2018fractional} for the sum-of-ratios maximization problem.
\begin{proposition}[Quadratic Transform \cite{shen2018fractional}]
The sum-of-ratios maximization problem in \eqref{prob:FP_max} is equivalent to\footnote{This equivalence does not require the concave-convex assumption.}
    \begin{align}
    \label{prob:FP_max:quadratic}  \underset{\bx\in\mathcal{X},\,\by\in\mathbb R^{N}}{\text{maximize}} &\quad \sum^N_{n=1}\left(2y_n\sqrt{A_n(\bx)}-y^2_nB_n(\bx)\right)
    \end{align}
in the sense that $\bx^\star$ is a solution to \eqref{prob:FP_max} if and only if $(\bx^\star,\by^{\star})$ is a solution to \eqref{prob:FP_max:quadratic}, where $\by=\{y_n\}_{n=1}^N$ represents a set of auxiliary variables.
\end{proposition}

The above problem rewriting greatly eases the optimization of $\bx$. Since each $A_n(\bx)$ is concave and each $B_n(\bx)$ is convex, optimizing $\bx$ for fixed $\by$ is a convex problem that can be efficiently solved by the standard method; when $\bx$ is held fixed, each auxiliary variable $y_n$ can be optimally updated as
\begin{equation}
\label{max_FP:y}
    y_n = \frac{\sqrt{A_n(\bx)}}{B_n(\bx)}.
\end{equation}
As shown in \cite{shen2018fractional}, the alternating optimization between $\bx$ and $\by$ increases the sum-of-ratios objective value monotonically after each iteration. { Furthermore, if the functions $A_n(\bx)$ and $B_n(\bx)$ are all differentiable, then the alternating optimization yields convergence to a stationary point.} Actually, the use of quadratic transform is not limited to the sum-of-ratios problem and is not even limited to continuous variables; we refer the interested readers to \cite{shen2018fractional} and \cite{shen2018fractional1}.

\subsection{Min-FP Case}
We now shift our attention to the convex-concave sum-of-ratios minimization problem in \eqref{prob:FP_min}. One may suggest rewriting \eqref{prob:FP_min} in the max-FP form as
\begin{align}
\underset{\bx\in\mathcal X}{\text{maximize}} &\quad \sum^N_{n=1}\frac{A_n(\bx)}{\big(-B_n(\bx)\big)}
  \label{prob:FP_min:minus}
\end{align}
thus applying the quadratic transform. But this approach is problematic because \eqref{prob:FP_min:minus} violates the requirement that every denominator is strictly positive. Another natural idea is to flip each ratio and then turn \eqref{prob:FP_min} to
\begin{align}
\underset{\bx\in\mathcal X}{\text{maximize}} &\quad \sum^N_{n=1}\frac{B_n(\bx)}{A_n(\bx)}.
  \label{prob:FP_min:flip}
\end{align}
Notice that \eqref{prob:FP_min} and \eqref{prob:FP_min:flip} are not equivalent when $N\ge2$. Actually, solving \eqref{prob:FP_min:flip} boils down to minimizing a lower bound on the original objective function in \eqref{prob:FP_min}; this can be seen by recognizing $(\sum^N_{n=1}{A_n}/{B_n})/N$ as an arithmetic mean which is bounded below by the harmonic mean $N/(\sum^N_{n=1} {B_n}/{A_n})$. %This approximation can be fairly loose.

The following proposition provides the right way to extend the quadratic transform to the min-FP case.

\begin{proposition}[Inverse Quadratic Transform]
\label{proposition:IQT}
The sum-of-ratios minimization problem in \eqref{prob:FP_min} is equivalent to    
{\begin{align}
\label{prob:FP_min:quadratic}
\underset{\bx\in\mathcal{X}, \,\tilde\by\in\mathbb{R}^N}{\text{minimize}} &\quad \sum^N_{n=1}\frac{1}{[2\tilde y_n\sqrt{B_n(\bx)}-\tilde y_n^2A_n(\bx)]_+}
\end{align}}in the sense that $\bx^\star$ is a solution to \eqref{prob:FP_min} if and only if $(\bx^\star,\tilde\by^\star)$ is a solution to \eqref{prob:FP_min:quadratic}, where $\tilde\by=\{\tilde y_n\}_{n=1}^N$ represents a set of auxiliary variables.
\end{proposition}
\begin{IEEEproof}
When $\bx$ is fixed, each $\tilde y_n$ in  \eqref{prob:FP_min:quadratic} is optimally determined as $\tilde y^\star_n=\sqrt{B_n(\bx)}/A_n(\bx)$. 
Plugging this $\tilde y^\star_n$ into the new objective function \eqref{prob:FP_min:quadratic} recovers the original objective function in \eqref{prob:FP_min}. The equivalence is then established.
\end{IEEEproof}

% {The above transformation incurs an additional operation $\text{max}\{\cdot,0^+\}$.} 
{Putting $2\tilde y_n\sqrt{B_n(\bx)}-\tilde y_n^2A_n(\bx)$ in $[\cdot]_+$ is necessary; otherwise, it would be optimal to let $y_n\rightarrow0_-$ and consequently \eqref{prob:FP_min} and \eqref{prob:FP_min:quadratic} are no longer equivalent.}

\begin{remark}
    Under the convex-concave assumption, i.e., if each $A_n(\bx)$ is a convex function and each $B_n(\bx)$ is a concave function, the new problem \eqref{prob:FP_min:quadratic} is convex in $\bx$ for fixed $\tilde \by$.
\end{remark}

We then obtain an efficient algorithm for addressing  \eqref{prob:FP_min}: in an alternating fashion, update $\bx$ by solving the convex problem in \eqref{prob:FP_min:quadratic} with fixed $\by$, and then update $\by$ as
\begin{equation}
\label{min_FP:y}
    \tilde y_n = \frac{\sqrt{B_n(\bx)}}{A_n(\bx)+\varepsilon},
\end{equation}
where a small positive constant $\varepsilon>0$ is needed to prevent $\tilde y_n$ from growing to infinity as $A_n(\bx)\rightarrow0_+$.

We further state a generic form of inverse quadratic transform in the following proposition. Its proof follows that of Proposition \ref{proposition:IQT} closely and is omitted here.

%In fact, the inverse quadratic transform can be further extended to a sum-of-functions-of-ratios problem, a more general cost-function-of-multiple-ratios problem, and a special min-max-ratios problem, as specified in the following.

\begin{proposition}
\label{theorem:sum of functions of ratios}
Assuming that each ratio $A_n(\bx)/B_n(\bx)$ in \eqref{prob:FP_min} is now the argument of a convex nondecreasing function $f_n:\mathbb R_+\rightarrow\mathbb R$, the \emph{sum-of-functions-of-ratios} problem
\begin{align}
\underset{\bx\in\mathcal{X}}{\text{minimize}} &\quad \sum^N_{n=1}f_n\left(\frac{A_n(\bx)}{B_n(\bx)}\right)
   \label{prob:FP_min_fun}
\end{align}
is equivalent to 
    % \begin{align}
    % \label{prob:FP_min_fun:quadratic}
    %      \underset{\bx\in\mathcal{X},\,\tilde \by\in\mathbb R^N}{\text{minimize}} &\quad \sum^N_{n=1}f_n\left(\frac{1}{{\text{max}\{2\tilde y_n\sqrt{B_n(\bx)}-\tilde y_n^2A_n(\bx), 0^+\}}}\right)
    % \end{align}
{\begin{align}
\label{prob:FP_min_fun:quadratic}
\underset{\bx\in\mathcal{X},\,\tilde \by\in\mathbb R^N}{\text{minimize}} &\quad \sum^N_{n=1}f_n\left(\frac{1}{[2\tilde y_n\sqrt{B_n(\bx)}-\tilde y_n^2A_n(\bx)]_+}\right)
\end{align}}in the sense that $\bx^{\star}$ is a solution to \eqref{prob:FP_min_fun} if and only if $(\bx^{\star},\tilde \by^{\star})$ is a solution to \eqref{prob:FP_min_fun:quadratic}.
\end{proposition}

The alternating optimization between $\bx$ and $\tilde \by$ is still easy to carry out, i.e., with each $\tilde y_n$ iteratively updated in closed form as in \eqref{min_FP:y}, solving for $\bx$ in \eqref{prob:FP_min_fun:quadratic} is a convex optimization problem. {The performance of this alternating optimization is analyzed in Section \ref{subsec:MM & convergence} by means of MM.}

\subsection{Mixed Max-and-Min FP Case}
We here propose a new type of FP problem that incorporates max-FP and min-FP. Still, consider a total of $N$ ratios. Let each $f^+_n:\mathbb{R}_{+}\rightarrow\mathbb R$ be a concave increasing function, for $n=1,\ldots,N_0$; let each $f^-_n:\mathbb R_{+}\rightarrow\mathbb R$ be a concave decreasing function, for $n=N_0+1,\ldots,N$. The resulting \emph{mixed max-and-min FP} problem is defined to be
\begin{align}
\label{prob:FP_mixed}
\underset{\bx\in\mathcal X}{\text{maximize}} &\, \sum_{n\le N_0}f^+_n\left(\frac{A_n(\bx)}{B_n(\bx)}\right)+\sum_{n>N_0}f^-_n\left(\frac{A_n(\bx)}{B_n(\bx)}\right).
\end{align}
Intuitively, we aim to maximize the ratios inside $f^+_n(\cdot)$ while minimizing those inside $f^-_n(\cdot)$.
% \footnote{This difference lies in the different monotonicity of $f^+_n(\cdot)$ and $f^-_n(\cdot)$}.
Again, we let each $A_n(\bx)$ be concave and each $B_n(\bx)$ be convex in the max-FP part, and reverse the concavity and convexity in the min-FP part.

As the building block of this paper, the quadratic transform in \cite{shen2018fractional} and the inverse quadratic transform in Proposition \ref{proposition:IQT} can be integrated into a unified approach to the above mixed FP problem, as stated in the following theorem.

\begin{theorem}[Unified Quadratic Transform]
\label{theorem:UQT}
    The mixed max-and-min FP problem in \eqref{prob:FP_mixed} is equivalent to
{\begin{subequations}
\label{prob:FP_mixed:quadratic}
\begin{align}
\underset{\bx\in\mathcal{X},\by,\tilde{\by}}{\text{maximize}} &\;\; \sum_{n\le N_0}f^+_n\left(2y_n\sqrt{A_n(\bx)}-y^2_nB_n(\bx)\right)\,+\notag\\
&\;\;\;\sum_{n>N_0}f^-_n\left(\frac{1}{[2\tilde y_n\sqrt{B_n(\bx)}-\tilde y_n^2A_n(\bx)]_+}\right)
\label{obj:FP_mixed_fun:quadratic}\\
\text{subject to}
&\;\;\; y_n,\tilde{y}_n\in\mathbb R,
\label{prob:FP_mixed_fun:quadratic:cons_c}
\end{align}    
\end{subequations}}in the sense that $\bx^\star$ is a solution to \eqref{prob:FP_mixed} if and only if $(\bx^\star,\by^\star,\tilde{\by}^\star)$ is a solution to \eqref{prob:FP_mixed:quadratic}, where $\by=\{y_n\}_{n=1}^{N_0}$ is a set of auxiliary variables for max-FP while $\tilde\by=\{\tilde y_n\}_{n=N_0+1}^{N}$ is a set of auxiliary variables for min-FP.
\end{theorem}

The above result can be readily verified by putting together the proof of quadratic transform in \cite{shen2018fractional} and the proof of Proposition \ref{proposition:IQT}. Again, an alternating optimization can be carried out based on the above problem transformation. When $\bx$ is fixed, each $y_n$ is updated according to \eqref{max_FP:y} and each $\tilde{y}_n$ is updated according to \eqref{min_FP:y}; when $\by$ and $\tilde{\by}$ are fixed, $\bx$ in \eqref{prob:FP_mixed:quadratic} can be optimally determined via convex optimization. 

{We will further show in Section \ref{sec:MM} that the alternating optimization between $\bx$ and $\{\by,\tilde\by\}$ in \eqref{prob:FP_mixed:quadratic} converges to a stationary point. The main idea is to interpret this alternating optimization as an MM method\footnote{{The MM method is referred to as the successive upper-bound minimization (SUM) method in \cite{razaviyayn2013unified} with certain conditions being satisfied.}}---which is a specific case of the inexact block coordinate descent (BCD) \cite{razaviyayn2013unified} in the sense that it uses an upper bound to approximate the objective function successively. As compared to the BCD method that requires each block optimization subproblem to be uniquely solvable in order to attain a stationary point \cite{bertsekas1999nonlinear}, the inexact BCD method can drop this uniqueness requirement \cite{razaviyayn2013unified}.}

\begin{figure}[t]
  \centering
  \includegraphics[width=8cm]{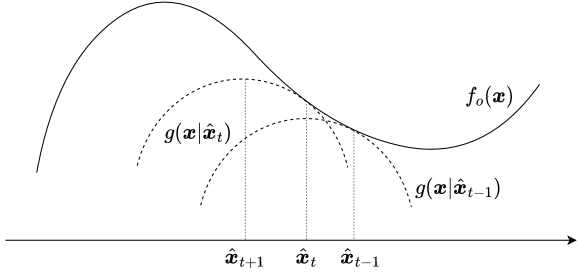}
  \caption{The MM procedure. The original objective function is denoted by $f_o(\bx)$; the subscript $t$ is the iteration index. The proposed FP method can be interpreted as constructing surrogate function $g(\bx|\hat\bx)$ successively.}
  \label{fig:MM}
\end{figure}

{\subsection{A Multi-Objective Optimization Perspective}
We now compare the different FP methods from a multi-objective optimization perspective.
Consider a sequence of FP problems each with a single ratio to maximize, i.e., 
\begin{align*}
\underset{\bx\in\mathcal X}{\text{maximize}}&\quad \frac{A_n(\bx)}{B_n(\bx)},\quad \text{for}\; n=1,\ldots,N_0.
\end{align*}
 A typical multi-objective problem is
\begin{align*}
\underset{\bx\in\mathcal X}{\text{maximize}}&\quad\min_n\;\frac{A_n(\bx)}{B_n(\bx)}.
\end{align*}
As shown in \cite{crouzeix1985algorithm}, a generalized Dinkelbach's algorithm suffices to address the above problem. Nevertheless, if we consider a linear combination of these single-ratio objectives, i.e.,
\begin{align*}
\underset{\bx\in\mathcal X}{\text{maximize}}&\quad\sum^{N_0}_{n=1}\alpha_n\cdot\frac{A_n(\bx)}{B_n(\bx)},
\end{align*}
where each weight $\alpha_n\ge0$. Then the generalized Dinkelbach's algorithm in \cite{crouzeix1985algorithm} no longer works, but the existing quadratic transform in \cite{shen2018fractional} is applicable.}

{Furthermore, if we in addition consider a sequence of FP problems each with a single ratio to minimize, i.e., 
\begin{align*}
\underset{\bx\in\mathcal X}{\text{minimize}}&\quad \frac{A_n(\bx)}{B_n(\bx)}, \quad \text{for} \;n=N_0+1,\ldots,N.
\end{align*}
The max-FP and the min-FP problems can be combined into
\begin{align*}
\underset{\bx\in\mathcal X}{\text{maximize}}&\quad\sum^{N_0}_{n=1}\alpha_n\cdot\frac{A_n(\bx)}{B_n(\bx)}-\sum^{N}_{n=N_0+1}\alpha_n\cdot\frac{A_n(\bx)}{B_n(\bx)},
\end{align*}
where each weight $\alpha_n\ge0$. Now even the quadratic transform in \cite{shen2018fractional} does not work. The proposed unified quadratic transform in Theorem \ref{theorem:UQT} then comes into play.}

%Next, we draw further insights into this unified method by the MM theory.

\section{Connection to Minorization-Maximization}
\label{sec:MM}

\subsection{MM Interpretation of Proposed Unified Approach}
\label{subsec:MM & convergence}

{
Let us first review the MM method briefly. We adopt the notation in \cite{shen2019optimization,sun2016majorization}. For the original problem
\begin{align*}
    \underset{\bx\in\mathcal X}{\text{maximize}}& \quad f(\bx),
\end{align*}
we construct a \emph{surrogate function} $g(\bx|\hat\bx)$ of $\bx$ given the condition parameter $\hat\bx\in\mathcal X$, such that
$$
g(\bx|\hat\bx)\le f(\bx) \quad\text{and}\quad
g(\hat\bx|\hat\bx)= f(\hat\bx).
$$
Then instead of optimizing $\bx$ directly in the original problem, the MM method solves a sequence of new problems of the surrogate function:
\begin{align*}
    \underset{{\bx\in\mathcal X}}{\text{maximize}}& \quad g(\bx|\hat\bx),
\end{align*}
with $\hat\bx$ iteratively updated to the previous solution $\bx$, as illustrated in Fig.~\ref{fig:MM}.}

It turns out that Theorem \ref{theorem:UQT} amounts to constructing a surrogate function for \eqref{prob:FP_mixed}. The optimal updates of $\by$ and $\tilde\by$ in \eqref{max_FP:y} and \eqref{min_FP:y} are treated as two functions of $\bx$:
\begin{equation}
    Y_n(\bx) = \frac{\sqrt{A_n(\bx)}}{B_n(\bx)}\quad\text{and}\quad \tilde{Y}_n(\bx) = \frac{\sqrt{B_n(\bx)}}{A_n(\bx)}.
\end{equation}
Denoting by $f_o(\bx)$ the objective function in \eqref{prob:FP_mixed}, we claim that 
\begin{align}
\label{g}
&g(\bx|\hat\bx)=\sum_{n=1}^{N_0}f^+_n\left(2Y_n(\hat{\bx})\sqrt{A_n(\bx)}-(Y_n(\hat{\bx}))^2B_n(\bx)\right)\,+\notag\\
&\sum_{n=N_0+1}^{N}f^-_n\bigg(\frac{1}{[{2\tilde{Y}_n(\hat{\bx})\sqrt{B_n(\bx)}-(\tilde{Y}_n(\hat{\bx}))^2A_n(\bx)}]_+}\bigg)
\end{align}
is a surrogate function. {To show that, we first let $\bx=\hat\bx$ and then have
$$
g(\hat\bx|\hat\bx) = \sum_{n=1}^{N_0}f^+_n\left(\frac{A_n(\hat\bx)}{B_n(\hat\bx)}\right)+\sum_{n=N_0+1}^{N}f^-_n\left(\frac{A_n(\hat\bx)}{B_n(\hat\bx)}\right)=f_o(\hat\bx).
$$
Furthermore, defining
\begin{align}
{\Theta}_n&=Y_n(\hat\bx)\sqrt{B_n(\bx)}-\sqrt{A_n(\bx)}/ \sqrt{B_n(\bx)},\\
\tilde{\Theta}_n &= \tilde {Y}_n(\hat\bx)\sqrt{A_n(\bx)}-\sqrt{B_n(\bx)}/ \sqrt{A_n(\bx)},
\end{align}
we have
\begin{align}   &\sum_{n=1}^{N_0}f^+_n\left(2Y_n(\hat{\bx})\sqrt{A_n(\bx)}-(Y_n(\hat{\bx}))^2B_n(\bx)\right)\notag\\
&=\sum_{n=1}^{N_0}f_n^{+}\bigg(\frac{A_n(\bx)}{B_n(\bx)}-{\Theta}_n^2\bigg)\notag\\
&\le  \sum_{n=1}^{N_0}f_n^{+}\bigg(\frac{A_n(\bx)}{B_n(\bx)}\bigg),
\end{align}
where the inequality follows since $f^+_n(\cdot)$ is an increasing function, and also
\begin{align}   &\sum_{n=N_0+1}^{N}f^-_n\bigg(\frac{1}{[{2\tilde{Y}_n(\hat{\bx})\sqrt{B_n(\bx)}-(\tilde{Y}_n(\hat{\bx}))^2A_n(\bx)}]_+}\bigg)\notag\\
&=\sum_{n=N_0+1}^{N}f_n^{-}\bigg(\frac{1}{[B_n(\bx)/A_n(\bx)-\tilde{\Theta}_n^2]_+}\bigg)\notag\\
&\le\sum_{n=N_0+1}^{N}f_n^{-}\bigg(\frac{A_n(\bx)}{B_n(\bx)}\bigg),
\end{align}
where the inequality can be shown as follows: (i) if $B_n(\bx)/A_n(\bx)\ge\tilde\Theta^2_n$, then the inequality holds since $f^-_n(\cdot)$ is a decreasing function; (ii) if $B_n(\bx)/A_n(\bx)<\tilde\Theta^2_n$, then we have $f^-_n\big(\frac{1}{[B_n(\bx)/A_n(\bx)-\tilde{\Theta}_n^2]_+}\big)=f_n^-(\infty)\le f^-_n(\frac{A_n(\bx)}{B_n(\bx)}) $.}

Combining the above two inequalities gives $g(\bx|\hat\bx)\le f_o(\bx)$. Thus, $g(\bx|\hat\bx)$ constitutes a surrogate function for $f_o(\bx)$, and hence
the alternating optimization between $\bx$ and $\{\by,\tilde\by\}$ in \eqref{prob:FP_mixed:quadratic} can be recognized as an MM method. In light of the MM theory \cite{razaviyayn2013unified}, we immediately have the following result.

\begin{proposition}[Convergence Analysis Based on \cite{razaviyayn2013unified}]
\label{prop:convergence}
    % {The $g(\bx|\hat\bx)$ in \eqref{g} is a surrogate function of the mixed max-and-min objective function in \eqref{prob:FP_mixed}.}
    The alternating optimization between $\bx$ and $(\by,\tilde\by)$ in \eqref{prob:FP_mixed:quadratic} yields a nondecreasing convergence of the mixed max-and-min FP objective in \eqref{prob:FP_mixed}. {Furthermore, if the functions $f^+_n(\cdot)$, $,f^-_n(\cdot)$, $A_n(\cdot)$, and $B_n(\cdot)$ are all differentiable, then the alternating optimization converges to a stationary point of \eqref{prob:FP_mixed}.}
\end{proposition}

%Clearly, the above analysis readily carries over to the inverse quadratic transform in Proposition \ref{proposition:IQT} since it is a special case of the unified quadratic transform. 

{\begin{remark}
We could have rewritten problem \eqref{prob:FP_min:quadratic} as
\begin{subequations}
\begin{align}
\underset{\bx\in\mathcal{X}, \,\tilde\by\in\mathbb{R}^N}{\text{minimize}} &\quad \sum^N_{n=1}\frac{1}{2\tilde y_n\sqrt{B_n(\bx)}-\tilde y_n^2A_n(\bx)}\\
\text{subject to}&\quad\; 2\tilde y_n\sqrt{B_n(\bx)}-\tilde y_n^2A_n(\bx) > 0,\quad\forall n.
\label{rewriting:constraint}
\end{align}    
\end{subequations}
However, the constraint \eqref{rewriting:constraint} is troublesome in the convergence analysis since it causes coupling between $\bx$ and $\tilde\by$ in the alternating optimization. For this reason, we incorporate this constraint into the objective function by using $[\cdot]_+$ in the previous discussion. 
\end{remark}}

\subsection{Matrix Extension}

Thus far our discussion is restricted to the traditional FP scenario with scalar-valued numerators and denominators. However, the practical problems in wireless networks may involve ratios between matrices.
For example, the SINR is a matrix ratio when we consider multi-input multiple-output (MIMO) transmission, and so is the Cram\'{e}r-Rao bound when we consider radar sensing. 
We are thus motivated to study the matrix extension of the unified quadratic transform.

First, let us extend the mixed max-and-min FP problem \eqref{prob:FP_mixed} to the matrix ratios. Assume that each $\bA_n(\bx)\in\mathbb H^{d\times d}_{+}$ and each $\bB_n(\bx)\in\mathbb H^{d\times d}_{++}$. We denote by $\sqrt{\bM}\in\mathbb C^{d\times \ell}$ the square root of $\bM$ for some positive integer $\ell\le d$, i.e., $\sqrt{\bM}(\sqrt{\bM})^{\text{H}}=\bM$. Let $f^+_n:\mathbb H_{+}^{\ell\times\ell}\rightarrow\mathbb R$ be a concave increasing function so that $f^+_n(\bM)>f_n^+(\bM')$ if $\bM\succ\bM'$, $n=1,\ldots,N_0$. Similarly, let $f^-_n:\mathbb H_{+}^{\ell\times\ell}\rightarrow\mathbb R$ be a concave decreasing function so that $f^-_n(\bM)<f^-_n(\bM')$ if $\bM\succ\bM'$, $n=N_0+1,\ldots,N$. In particular, we require the following cyclic property for every min-FP function $f^-_n(\cdot)$:
\begin{equation}
\label{property:f^-}
f^-_n((\sqrt{\bA}^{\text{H}}\bB^{-1}\sqrt{\bA})^{-1}) = f^-_n(\sqrt{\bB}^{\text{H}}\bA^{-1}\sqrt{\bB}),
\end{equation}
e.g., $f^-_n(\bm X)=\mathrm{tr}(\bm X)$ and $f^-_n(\bm X)=\log\det(\bI+\bm X)$ both meet the above condition. The reason for assuming \eqref{property:f^-} can be seen in the proof of Theorem \ref{theorem:matrix_surrogate}.

We propose the following matrix extension of problem \eqref{prob:FP_mixed}:
\begin{align}
\label{prob:FP_mixed_matrix}
 \underset{\bx\in\mathcal X}{\text{maximize}} &\;\sum_{n=1}^{N_0}f^+_n\left(\sqrt{\bA}^{\text{H}}_n(\bx)\bB^{-1}_n(\bx)\sqrt{\bA}_n(\bx)\right)\notag\\
&\;+\sum_{n=N_0+1}^{N}f^-_n\Big(\sqrt{\bA}^{\text{H}}_n(\bx)\bB^{-1}_n(\bx)\sqrt{\bA}_n(\bx)\Big).
\end{align}
Inspired by the scalar case in \eqref{g}, we suggest constructing a surrogate function for the above matrix-ratio objective as 
\begin{align}
\label{matrix_g}
g(\bx|\hat\bx)=\sum_{n=1}^{N_0}f^+_n\left(\bQ_n^+(\bx|\hat\bx)\right)
+\sum_{n=N_0+1}^{N}f^-_n\left(({\bQ_n^-}(\bx|\hat\bx))^{-1}\right)
\end{align}
with the shorthand 
\begin{multline}
    \bQ^+_n(\bx|\hat\bx)=\sqrt{\bA}^{\text{H}}_n(\bx) \bY_n(\hat\bx) + \bY^{\text{H}}_n(\hat\bx)\sqrt{\bA}_n(\bx)\\
    -\bY_n^{\text{H}}(\hat\bx) \bB_n(\bx) \bY_n(\hat\bx),
     \label{matrix_+_g}
\end{multline}
and
\begin{multline}
\bQ^-_n(\bx|\hat\bx)=\sqrt{\bB}^{\text{H}}_n(\bx) \tilde{\bY}_n(\hat\bx) + \tilde\bY^{\text{H}}_n(\hat\bx)\sqrt{\bB}_n(\bx)\\
-\tilde{\bY}_n^{\text{H}}(\hat\bx) \bA_n(\bx) \tilde{\bY}_n(\hat\bx),
    \label{matrix_-_g}
\end{multline}
where the matrix functions $\bY_n(\bx)$ and 
$\tilde\bY_n(\bx)$ are defined as 
\begin{align}
    &\bY_n(\bx) = \bB_n^{-1}(\bx)\sqrt{\bA_n}(\bx),
    \label{eq:matrix_y_function:+}\\
    &\tilde{\bY}_n(\bx) = \bA_n^{-1}(\bx)\sqrt{\bB_n}(\bx).
    \label{eq:matrix_y_function:-}
\end{align}

\begin{theorem}
\label{theorem:matrix_surrogate}
    The function $g(\bx|\hat\bx)$ in \eqref{matrix_g} is a surrogate function of the matrix mixed max-and-min objective function in \eqref{prob:FP_mixed_matrix}, so problem \eqref{prob:FP_mixed_matrix}
is equivalent to
\begin{subequations}
  \label{prob:FP_mixed_matrix:quadratic}
\begin{align}
\underset{\bx\in\mathcal X,\,\bY_n,\,\tilde{\bY}_n}{\text{maximize}}&\; \sum_{n=1}^{N_0}f^+_n\left(\bQ^+_n\right)+\sum_{n=N_0+1}^{N}f^-_n\left((\bQ^-_n)^{-1}\right)
\label{prob:FP_mixed_matrix_fun:quadratic:obj}\\
\text{subject to}
&\quad\;\bQ^-_n \succ \mathbf{0},\,\forall n>N_0
\label{prob:FP_mixed_matrix_fun:quadratic:cons_c}\\
&\quad\;\bY_n, \tilde\bY_n\in\mathbb C^{d\times\ell},
\end{align}    
\end{subequations}
where
\begin{align}
    \bQ^+_n&={\sqrt{\bA}^{\text{H}}_n(\bx) \bY_n + \bY^{\text{H}}_n\sqrt{\bA}_n(\bx)- \bY_n^{\text{H}} \bB_n(\bx) \bY_n},\\
    \bQ^-_n&=\sqrt{\bB}^{\text{H}}_n(\bx) \tilde{\bY}_n + \tilde{\bY}^{\text{H}}_n\sqrt{\bB}_n(\bx)- \tilde{\bY}_n^{\text{H}} \bA_n(\bx) \tilde{\bY}_n,
\end{align}
in the sense that $\bx^\star$ is a solution to \eqref{prob:FP_mixed_matrix} if and only if $(\bx^\star,\{\bY_n^\star\}^{N_0}_{n=1},\{\tilde{\bY}_n^\star\}^{N}_{n=N_0+1})$ is a solution to \eqref{prob:FP_mixed_matrix:quadratic}.
\end{theorem}
\begin{IEEEproof}
    See Appendix.
\end{IEEEproof}

The alternating optimization between $\bx$ and $(\{\bY_n\},\{\tilde{\bY}_n\})$ follows. When $\bx$ is held fixed, $(\{\bY_n\},\{\tilde{\bY}_n\})$ are optimally updated according to \eqref{eq:matrix_y_function:+} and \eqref{eq:matrix_y_function:-}. Subsequently, after the auxiliary variables have been updated, solving for $\bx$ in \eqref{prob:FP_mixed_matrix:quadratic} is a convex problem. By the MM theory, the convergence results in Proposition \ref{prop:convergence} continue to hold for matrix ratios, as stated in the following proposition.

\begin{proposition}
\label{prop:FP_mixed_matrix}
    The alternating optimization between $\bx$ and $(\{\bY_n\},\{\tilde{\bY}_n\})$ in the new problem \eqref{prob:FP_mixed_matrix:quadratic} yields a nondecreasing convergence of the original matrix-FP objective value in \eqref{prob:FP_mixed_matrix}. {Furthermore, if functions $f^+_n(\cdot)$, $,f^-_n(\cdot)$, $\bA_n(\cdot)$, and $\bB_n(\cdot)$ are all differentiable, then the alternating optimization converges to a stationary point of the matrix-ratio FP problem \eqref{prob:FP_mixed_matrix}.}
\end{proposition}

\subsection{Generalized Lagrangian Dual Transform}
\label{subsec:Lagrangian}

We now propose a new technique that can further facilitate the solving of the mixed max-and-min problem \eqref{prob:FP_mixed} when the ratios reside in logarithms---which can be frequently countered in wireless communications because of Shannon's formula $C=\log(1+\mathrm{SINR})$. Indeed, the unified quadratic transform in Theorem \ref{theorem:UQT} alone can already address this type of problem, i.e., by treating each SINR as a ratio inside the increasing concave function $f^+(r)=\log(1+r)$. The log-ratio problem is then recast to a sequence of convex problems with the auxiliary variables iteratively updated.  However, ``log'' may still cause the optimization solver (e.g., CVX \cite{grant2020cvx}) to slow down or even fail to work\footnote{The authors of CVX warned in  \cite{grant2020cvx} that their software is slower and less reliable when dealing with models involving ``log" or other functions in the log, exp, and entropy family.

{We further look into the software implementation of CVX to figure out the underlying reason. Whenever the CVX solves a convex problem with log, the message box displays ``SDPT3 will be called several times to refine the solution.'' This is also the case for SeDuMi. The rationale is that the solver needs to approximate log successively.}}.

So how do we get rid of log if ratios are nested in it? An existing technique called \emph{Lagrangian dual transform} exactly serves this purpose \cite{shen2018fractional1}. Intuitively speaking, its main idea is to somehow move the ratios to the outside of log, thereby converting the log-ratio problem to the sum-of-ratios problem in \eqref{prob:FP_max}. And yet this technique \cite{shen2018fractional1} only applies to the max-FP case. As another main result of this paper, we show that the Lagrangian dual transform can be extended to the mixed case as well.

With $f^+_n(r)=w_n\log(1+r)$ and $f^-_n(r)=-w_n\log(1+r)$, the mixed max-and-min problem in \eqref{prob:FP_mixed} specializes to
\begin{align}
\label{prob:weighted sum-of-logarithms}
\underset{\bx\in\mathcal{X}}{\text{maximize}}&\quad\sum_{n=1}^{N_0}w_n\operatorname{log}\left(1+\frac{A_n(\bx)}{B_n(\bx)}\right)\notag\\
&\qquad\qquad-\sum_{n=N_0+1}^{N}w_n\operatorname{log}\left(1+\frac{A_n(\bx)}{B_n(\bx)}\right).
    \end{align}
If considering the max-FP component alone (i.e., when $N_0=N$), then \cite{shen2018fractional1} shows by using the Lagrangian dual theory that the surrogate function for MM can be constructed as
\begin{multline}
    g^{+}(\bx|\hat\bx) = \sum_{n=1}^{N_0}w_n\log(1+\Gamma_n(\hat\bx))-\sum_{n=1}^{N_0}w_n\Gamma_n(\hat\bx)\\     \quad\quad\quad+\sum_{n=1}^{N_0}\frac{w_n(1+\Gamma_n(\hat\bx))A_n(\bx)}{A_n(\bx)+B_n(\bx)},
\end{multline}
where
\begin{equation}
    \Gamma_n(\hat\bx) = \frac{A_n(\hat\bx)}{B_n(\hat\bx)}.
\end{equation}
For the general case with $N_0<N$, we propose generalizing the above surrogate function as
\begin{align}
    \label{La_g}
    g(\bx|\hat{\bx})=g^+(\bx|\hat{\bx})+g^-(\bx|\hat{\bx}),
\end{align}
where
\begin{multline}
        g^{-}(\bx|\hat\bx) = \sum_{n=N_0+1}^{N}w_n\log(1-\tilde{\Gamma}_n(\hat\bx))+\sum_{n=N_0+1}^{N}w_n\tilde\Gamma_n(\hat\bx)\\
        \quad\quad\quad-\sum_{n=N_0+1}^{N}w_n(1-\tilde\Gamma_n(\hat\bx))\frac{A_n(\bx)}{B_n(\bx)}
\end{multline}
and
\begin{equation}
    \tilde\Gamma_n(\hat\bx) = \frac{A_n(\hat\bx)}{A_n(\hat\bx)+B_n(\hat\bx)}. 
\end{equation}
By the MM theory, the original problem \eqref{prob:weighted sum-of-logarithms} can be recast to a new problem based on \eqref{La_g} in which the ratios are moved to the outside of logarithms, as stated in the following theorem; the proof is already shown in the above discussion.
\begin{theorem}[Generalized Lagrangian Dual Transform]
\label{theorem:Lagrangian Dual Transform}
The log-ratio problem in \eqref{prob:weighted sum-of-logarithms} is equivalent to 
\begin{subequations}
\label{prob:weighted sum-of-logarithms:Lagrangian}
    \begin{align}    \underset{\bx\in\mathcal{X},\bm{\gamma},\tilde{\bm{\gamma}}}{\text{maximize}}&\quad \zeta^{+}(\bx,\bm{\gamma})+\zeta^{-}(\bx,\tilde{\bm{\gamma}})\\
        \text{subject to}&\quad \gamma_n, \tilde{{\gamma}}_n\in\mathbb{R},
    \end{align}
\end{subequations}
in the sense that $\bx^{\star}$ is a solution to \eqref{prob:weighted sum-of-logarithms} if and only if $(\bx^{\star},\bm{\gamma}^{\star},\tilde{\bm{\gamma}}^{\star})$ is a solution to \eqref{prob:weighted sum-of-logarithms:Lagrangian}, where $\bm{\gamma}=\{\gamma_n\}_{n=1}^{N_0}$ is a set of auxiliary variables for the ratios inside $f^+_n(r)=w_n\log(1+r)$ while $\tilde{\bm\gamma}=\{\tilde\gamma_n\}_{n=N_0+1}^{N}$ is a set of auxiliary variables for the ratios inside $f^-_n(r)=-w_n\log(1+r)$, along with
\begin{multline}
    \zeta^{+}(\bx,\bm{\gamma}) = \sum_{n=1}^{N_0}w_n\log(1+\gamma_n)-\sum_{n=1}^{N_0}w_n\gamma_n\\ +\sum_{n=1}^{N_0}\frac{w_n(1+\gamma_n)A_n(\bx)}{A_n(\bx)+B_n(\bx)},
\end{multline}
and
\begin{multline}
    \zeta^{-}(\bx,\tilde{\bm{\gamma}}) = \sum_{n=N_0+1}^{N}w_n\log\left(1-\tilde\gamma_n\right)+\sum_{n=N_0+1}^{N}w_n\tilde\gamma_n\\
    -\sum_{n=N_0+1}^{N}w_n(1-\tilde\gamma_n)\frac{A_n(\bx)}{B_n(\bx)}.
\end{multline}
\end{theorem}

{Notice that the new problem still has logarithms, but these logarithms only contain the auxiliary variables $\{\bm\gamma,\tilde{\bm\gamma}\}$ that can be efficiently solved in closed form, as specified below.}

We propose an alternating optimization between $\bx$ and $(\bm\gamma,\tilde{\bm\gamma})$. In \eqref{prob:weighted sum-of-logarithms:Lagrangian}, when $\bx$ is held fixed, the auxiliary variables  can be optimally updated in closed form as
\begin{equation}
    \gamma_n = \frac{A_n(\bx)}{B_n(\bx)} \quad\text{and}\quad \tilde\gamma_n = \frac{A_n(\bx)}{A_n(\bx)+B_n(\bx)}.
    \label{gamma_updata:max}
\end{equation} 
Importantly, when $\bm\gamma$ and $\tilde{\bm\gamma}$ are fixed, the new problem \eqref{prob:weighted sum-of-logarithms:Lagrangian} can be recognized as a mixed max-and-min sum-of-ratios problem of $\bx$, to which the unified quadratic transform in Theorem \ref{theorem:UQT} applies. Although we could have applied the unified quadratic transform directly to the log-ratio problem, using the generalized Lagrangian dual transform to get rid of log can significantly boost the software solving in practice. 
% { We will show a practical use of Theorem \ref{theorem:Lagrangian Dual Transform} in Section \ref{sec:secure transmission}.}

\section{Applications}
\label{sec:app}

Three applications of our new FP method are discussed here, each representing a category of FP problems that cannot be efficiently addressed by the existing FP tools: (i) AoI minimization corresponds to the multi-ratio min-FP, (ii) the Cram\'{e}r-Rao bound minimization for sensing corresponds to the matrix min-FP, and (iii) the secure data rate maximization corresponds to the mixed max-and-min FP. The proposed algorithms all yield provable convergence to stationary points. {We remark that the default solver SDPT3 is adopted throughout our simulations; the other solvers such as SeDuMi work as well.}%Simulations 

%waveform design for sensing is a multi-matrix-ratio min-FP problem, and power control for secure transmission is a mixed max-and-min FP problem. According to the MM theory as stated in Propositions  \ref{prop:convergence} and \ref{prop:FP_mixed_matrix}, the proposed FP algorithms all guarantee convergence to a stationary point. 

\subsection{AoI Reduction}

\subsubsection{Background} The notion of AoI was introduced in the early 2010s \cite{kaul2011minimizing} to quantify the freshness of data packets, which is defined to be the time elapsed since the generation of the last successfully received update information from a source node. The rate control problem is to coordinate the data-generating rates across multiple source nodes in order to minimize the overall AoI. The pioneer work \cite{kaul2011minimizing} proposes a descent algorithm for minimizing AoI in a single-source system. Considering a homogeneous network with multiple sources so that all the sources have the same service rate, \cite{yates2018age,najm2018status} show that the rate control can be optimally determined in closed form. Rate control becomes much more difficult when the service rate varies from source to source, namely the heterogeneous network. The authors of \cite{huang2015optimizing} suggest a quasilinear approximation of the two-source heterogeneous network to facilitate rate control. %Some more recent works consider the AoI minimization task in various application scenarios, e.g., \cite{hatami2021aoi} proposes a reinforcement algorithm for the sensor networks, and \cite{hu2020aoi} proposes an ant-colony heuristic algorithm for the unmanned aerial vehicle (UAV)-assisted wireless networks. In contrast, this work focuses on the optimization aspect of the AoI problem.
To the best of our knowledge, this is the first work that views the AoI problem from an FP perspective.

\subsubsection{Problem Formulation}
\begin{figure}
\begin{minipage}[b]{1.0\linewidth}
      \centering
      \centerline{\includegraphics[width= 5.5cm]{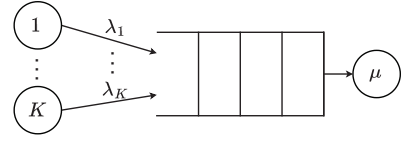}}
      %\vspace{-2em}
\end{minipage}
\caption{A multi-source queuing system with $K$ source nodes and a server. The data-generating rate of source $k$ is denoted by $\lambda_k$; the data processing rate at the server is denoted by $\mu$.}
\label{fig:diagram}
\end{figure}
\begin{figure}
\begin{minipage}[b]{1.0\linewidth}
      \centering
      \centerline{\includegraphics[width=6.5cm]{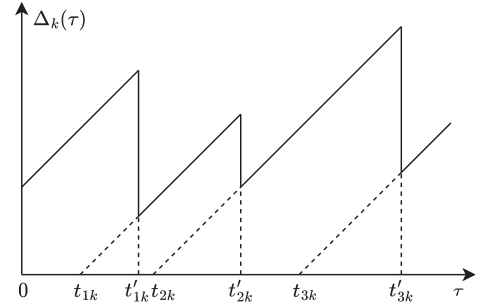}}
      % \vspace{-2em}
\end{minipage}
\caption{An example curve of AoI versus time. The average AoI amounts to the average area of the trapezoid below each tooth of the sawtooth curve.}
\label{fig:Age process}
\end{figure}

Consider a multi-source heterogeneous network wherein $K\ge2$ source nodes share a common server as depicted in Fig. \ref{fig:diagram}. Each source node $k=1,\ldots,K$ delivers data packets constantly toward the server at rate $\lambda_k$ (e.g., according to a Poisson process), while the processing rate at the server is $\mu$. The $i$th update packet from source $k$ is delivered at time $t_{ik}$ and finally departs the server at time $t'_{ik}$. The delay $t'_{ik}-t_{ik}$ is due to the queue waiting and the server processing. At the current time $\tau$, we use $\mathcal N_k(\tau)$ to denote the arrival time of the most recently received packet from source $k$:
\begin{equation}
    \mathcal N_k(\tau) = \max\left\{t'_{ik}:t'_{ik}\le \tau\right\}.
\end{equation}
The instantaneous AoI of source node $k$ at the present time $\tau$, denoted by $\Delta_k(\tau)$, is defined to be the time elapsed since its last update packet departs the server, i.e.,
\begin{equation}
    \Delta_k(\tau) = \tau - \mathcal N_k(\tau).
\end{equation}
As a result, $\Delta_k(\tau)$ increases linearly with $\tau$, and drops whenever a new update packet departs the server, so $\Delta_{k}(\tau)$ has a sawtooth profile along the time axis as illustrated in Fig. \ref{fig:Age process}. We are interested in the average AoI in the long run:
\begin{equation}
 \bar\Delta_k = \lim_{T\rightarrow\infty}\frac{1}{T}\int_{0}^{T}\Delta_k(\tau)d\tau,
\end{equation}
which can be recognized as the average area of the trapezoid below each tooth of the sawtooth curve as shown in Fig. \ref{fig:Age process}.

The specific form of $\bar\Delta_k$ depends on the queuing system model. For the M/M/1 model under LCFS-S, i.e., Last Come First Serve scheme that permits preempting package in Service with priority as considered in \cite{kaul2018age}, the average AoI of source $k$ is given by
\begin{equation}
\bar\Delta_{k}=\frac{1+\rho_{k}+3 \hat{\rho}_{k}+3 \hat{\rho}_{k} \rho_{k}+3 \hat{\rho}_{k}^{2}+\hat{\rho}_{k}^{2} \rho_{k}+\hat{\rho}_{k}^{3}}{\mu \rho_{k}\left(1+\hat{\rho}_{k}\right)},
\label{Eq_AoI_LCFS}
\end{equation}
where 
\begin{equation}
\rho_k= \frac{\lambda_k}{\mu}\quad\text{and}\quad\hat{\rho}_{k} = \sum_{i=1}^{k-1}\rho_i.
\end{equation}
Rate control aims to minimize the sum AoI by coordinating $\lambda_k$'s across source nodes, i.e.,
\begin{subequations}
    \label{prob:AoI}
\begin{align}
\underset{\bm\lambda}{\text{minimize}} &\quad \sum_{k=1}^K \bar\Delta_k    \label{prob:AoI:obj}\\
\text{subject to}
&\quad\; 0\le\lambda_k\le\mu,\;\text{for}\;k=1,\ldots,K.
\end{align}    
\end{subequations}
% which encompasses the traditional sum-of-AoI minimization  \cite{yates2018age,najm2018status,huang2015optimizing,hatami2021aoi,hu2020aoi} as a special case. The following proposed method also works for other types of AoI penalization, e.g., the sum-of-squared-AoI minimization.
\begin{algorithm}[t]
  \caption{Rate Control for Minimizing AoI}
  \label{algorithm:AoI}
  \begin{algorithmic}[1]
      \STATE Initialize $\bm{\lambda}$ to a feasible value.
      \REPEAT 
      \STATE Update each $\tilde{y}_k$ by \eqref{eq:AoI:y_update}.
      \STATE Solve the convex problem of $\mathbf{\lambda}$ in \eqref{prob:AoI:IQ}.
      \UNTIL{the objective value converges} 
  \end{algorithmic}
\end{algorithm}

\subsubsection{Proposed FP-Based Method}
The fractional term $\bar\Delta_{k}$ in \eqref{Eq_AoI_LCFS} strongly suggests the use of FP. But notice that $\bar\Delta_{k}$ does not meet the convex-concave assumption for min-FP. Nevertheless, this issue can be addressed by rewriting $\bar\Delta_k$ as a sum of two fractions
\begin{equation}
\bar\Delta_{k}=\frac{\hat{\rho}_{k}^{2}+3\hat{\rho}_{k}+1}{\mu(1+\hat{\rho}_k)}+\frac{(\hat{\rho}_k+1)^2}{\mu\rho_k},
\label{AoI_LCFS_equ}
\end{equation}
in which the numerators are all convex while the denominators are all concave. Accordingly, problem \eqref{prob:AoI} can be recast to the standard form in \eqref{prob:FP_mixed:quadratic} with $2K$ ratios:
\begin{subequations}
    \label{prob:AoI:final}
\begin{align}
\underset{\bm\lambda}{\text{maximize}} &\quad \sum_{k=1}^{K}f^{-}_k\left( \frac{\hat{\rho}_{k}^{2}+3\hat{\rho}_{k}+1}{\mu(1+\hat{\rho}_k)}\right)\notag\\
&\qquad\qquad\qquad+\sum_{k=K+1}^{2K}f_k^{-}\left(\frac{(\hat{\rho}_k+1)^2}{\mu\rho_k}\right)
\label{prob:AoI:equivalent:obj}\\
\text{subject to}
&\quad \,0\le\lambda_k\le\mu,\quad\forall k\\
&\quad \, Q_k>0,\quad\forall k
\end{align}    
\end{subequations}
where $f^-_k(r) = -r$ for all $k$. Notice that the above problem does not contain any max-FP components $f^+_k(\cdot)$. Next, by applying Theorem \ref{theorem:UQT}, we obtain the reformulation:
\begin{subequations}
\label{prob:AoI:IQ}
\begin{align}
    \underset{\bm{\lambda},\,\tilde{\bm{y}}}{\text{maximize}}&\quad \sum_{k=1}^K f_k^-((Q^-_k)^{-1})+\sum_{k=K+1}^{2K} f_{k}^{-}((Q^-_k)^{-1})\\
    \text{subject to}&\,\quad 0\le\lambda_k\le\mu,\quad\forall k
\end{align}
\end{subequations}
where
\begin{equation*}
  Q^-_{k} =
    \begin{cases}
      {2\tilde y_k\sqrt{\mu(1+\hat{\rho}_k)}-\tilde{y}_k^2(\hat{\rho}_{k}^{2}+3\hat{\rho}_{k}+1)} & \text{if $k\le K$,}\\
      {2\tilde{y}_k\sqrt{\mu\rho_k}-\tilde{y}_k^2(\hat{\rho}_k+1)^2} & \text{if $k>K$.}
    \end{cases}       
\end{equation*}
Now an iterative algorithm can be readily performed for the above new problem. When $\bm\lambda$ is held fixed, the auxiliary variables are optimally updated as
\begin{equation}
\tilde y_k =
    \begin{cases}
    {\sqrt{\mu(1+\hat{\rho}_k)}}/({\hat{\rho}_{k}^{2}+3\hat{\rho}_{k}+1+\varepsilon}) & \text{if $k\le K$,}\\
    {\sqrt{\mu\rho_k}}/\big({(\hat{\rho}_k+1)^2+\varepsilon}\big) & \text{if $k>K$.}
    \end{cases}  
    \label{eq:AoI:y_update}
\end{equation}
After the auxiliary variables have been updated, solving for $\bm\lambda$ in \eqref{prob:AoI:IQ} is a convex problem. Algorithm \ref{algorithm:AoI} summarizes the above steps.
% \begin{align}
%     c_{k}^{\star} = \,&\frac{\sqrt{\mu(1+\hat{\rho}_k)}}{\hat{\rho}_k^2+3\hat{\rho}_k+1},
%     \label{eq:AoI:c_update}\\
%     d_k^{\star} =\, &\frac{\sqrt{\mu\rho_k}}{(\hat{\rho}_k+1)^2}.
%     \label{eq:AoI:d_update}
% \end{align}
% Then, finding the optimal $\bm{\lambda}$ for fixed $\{\bm{c},\bm{d}\}$ is a convex problem. 

% This rate control method is summarized in Algorithm \ref{algorithm:AoI}. By Theorem \ref{theorem:convergence:UQT}, Algorithm \ref{algorithm:AoI} guarantees a convergence to a stationary point of the original problem \eqref{prob:AoI}.

%Notice that the objective remains an increasing convex function of these ratios.
\subsubsection{Numerical Results}

% \begin{figure}[t]
% \begin{minipage}[b]{1.0\linewidth}
% \centering
% \centerline{\includegraphics[width=8.5cm]{fig/convergence_N10.eps}} 
% \centerline{\small (a) Minimizing $\sum^K_{k=1}\bar\Delta_k$}
% \centerline{}
% \vspace{-0.5em}
% \end{minipage}
% \begin{minipage}[b]{1.0\linewidth}
% \centering
% \centerline{\includegraphics[width=8.5cm]{fig/quadratic_Iteration_N10.eps}}
% \centerline{\small (b) Minimizing $\sum^K_{k=1}\bar\Delta_k^2$}
% \end{minipage}
% \caption{\small Convergence of the proposed FP-based method.}
% \label{fig:converge}
% \end{figure}

\begin{figure}[t]
  \centering
  \includegraphics[width=8.5cm]{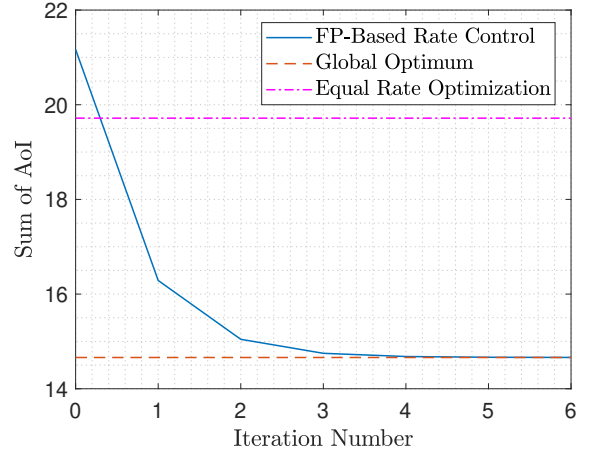}
  \caption{{Convergence of the FP-based rate control when $K=3$}.}
  \label{fig:converge}
\end{figure}

\begin{figure}[t]
  \centering
  \includegraphics[width=8.5cm]{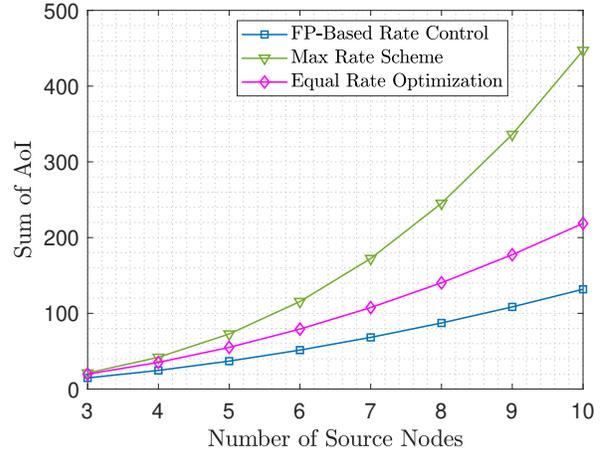}
  \caption{Sum of AoI incurred by the different rate control algorithms.}
  \label{fig:AoI different K}
\end{figure}

We test the performance of the proposed unified quadratic transform method by simulations. We fix the processing rate $\mu=1$ and try different numbers of source nodes $K=3,4,\ldots,10$. Consider two benchmarks: (i) \emph{Equal Rate Optimization} \cite{kaul2018age} that assumes all $\lambda_k$'s are equal and then performs the one-dimensional search; (ii) \emph{Max Rate Scheme} that sets each $\lambda_k=\mu$. {We also find the global optimum via an exhaustive search.} Observe from Fig.~\ref{fig:converge} that the proposed FP-based method has fast convergence; {it even converges to the global optimum in our simulation case}. Observe also that the majority of the AoI reduction is achieved by the proposed algorithm after just one iteration. Moreover, as shown in Fig.~\ref{fig:AoI different K}, the proposed method outperforms the benchmark methods significantly. For example, when $K=10$, the sum-of-AoI of the proposed method is approximately 40\% lower than that of the equal rate scheme \cite{kaul2018age}, and is approximately 70\% lower than that of the max rate scheme. 

% \begin{table}[t]
%   \centering
%   \small
%   \caption{Performance of the different rate control schemes}
%   \vspace{0.2em}
%     \begin{tabular}{c|cccc}
%     \hline
%     \hline
%      &$K=3$ & $K=5$ &$K=10$ & $K=20$  \\
%     \hline
%     Max Rate & $21.2$ & $72.7$ &$447.1$ & $3096.4$\\
%     Equal Rate \cite{kaul2018age} & $19.7$& $55.0$ & $218.8$ &$869.4$  \\
%     FP-Based Method &$14.7$ & $18.4$  & $131.8$ &$ 240.3$\\
%     \hline
%     \hline
%     \end{tabular}%
%   \label{tab:performance}%
% \end{table}

\subsection{Waveform Design for Multi-Radar Sensing}

\subsubsection{Background} A common goal of the waveform design for sensing is to minimize the Cram\'{e}r-Rao bound on the estimation error.
An early work in \cite{li2007range} shows that the Cram\'{e}r-Rao bound minimizing problem for the single-radar system is a convex problem under certain conditions. But the problem is nonconvex in most cases; existing optimization methods include successive convex approximation (SCA) \cite{cheng2019co} and semidefinite relaxation (SDR) \cite{liu2021cramer}. The multi-radar sensing problems are much more difficult. The authors of \cite{godrich2011sensor} suggest treating the multi-radar sensing problem as a knapsack problem and solving it by some heuristic algorithm. A three-stage method is proposed in the recent work \cite{yan2022target}: clutter suppression, tracklet detection, and tracklet association.% Another recent work \cite{zhang2022asynchronous} views the multi-radar sensing from an estimation fusion perspective and provides two asynchronous fusion algorithms. %Again, to the best of our knowledge, we are not aware of any existing FP-based studies in this area.
% Fortunately, our proposed unified quadratic transform works for it.

\subsubsection{Problem Formulation}
\begin{figure}[t]
  \centering
  \includegraphics[width=0.35\textwidth]{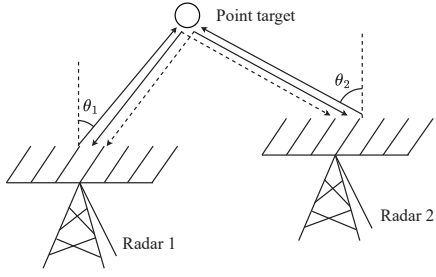}
  \caption{Two radar sets detect a common point target in the same spectrum band. The dashed lines represent mutual interference.}
  \label{fig:CRB_model}
\end{figure}
Consider $M$ radar sets that work on the same spectrum band to detect a common point target. Assume that the radar set $m$ has $N^T_m$ transmit antennas and $N^R_m$ receive antennas, both arranged as a half-wavelength spaced uniform linear array (ULA). Let $L$ be the number of samples taken of the echo signal at each radar set. For each radar $m$, denote by $\theta_m$ the direction of arrival (DOA) of the target as illustrated in Fig.~\ref{fig:CRB_model}, $\bS_m\in\mathbb{C}^{N_m^T\times L}$ the transmit waveform matrix, $\ba^{T}_{m}(\theta_m)\in\mathbb{C}^{N^{T}_{m}\times 1}$ the steering vector of the transmit antennas, $\mathbf{a}^{R}_{m}(\theta_m)\in\mathbb{C}^{N^{R}_{m}\times 1}$ the steering vector of the receive antennas, and $\bZ_m\sim \mathcal{CN}(\mathbf 0,\sigma_m^2\bI_{N^R_m})$ the background noise. Moreover, for a pair of radar sets $m$ and $m'$, let $\beta_{mm'}$ be the reflection coefficient from radar $m'$ to radar $m$ (which depends on the path loss and the radar cross section). Thus, the sampled echo signal $\bF_{m}\in\mathbb{C}^{N^{R}_{m}\times L}$ at the radar set $m$ is given by
\begin{equation}
\label{eq:double radar}
\bF_{m}=\sum_{m'=1}^{M}\beta_{mm'}\ba^{R}_{m}(\theta_m)(\ba^{T}_{m'}(\theta_{m'}))^{\top}\bS_{m'}+\bZ_m.
\end{equation}
Each radar set $m$ aims to estimate the DoA $\theta_m$ based on $\bF_m$. We seek the optimal waveform design of each $\bS_m$ under the power constraint $P_m$, i.e., with the squared Frobenius norm $\|\bS_m\|^2_F\le P_m$, in order to minimize the sum mean squared error (MSE) of all the $\theta_m$ estimates.

But the MSE of each $\theta_m$ is difficult to evaluate directly. A common practice is to use the Cram\'{e}r-Rao bound to approximate the actual MSE, as discussed in the sequel. Define the response matrix
\begin{equation}
\label{eq:target response matrix}
\bG_{mm'} = \beta_{mm'}\ba^{R}_{m}(\theta_m)(\ba^{T}_{m'}(\theta_{m'}))^{\top},
\end{equation} 
where the transmit steering vector $ \ba^{T}_{m}(\theta_m)$ and receive steering vector  of $\ba^{R}_{m}(\theta_m)$ radar $m$ can be written as 
\begin{align}
    \ba^{T}_{m}(\theta_m)=&\left[1,e^{-j\pi\sin\theta_m},\ldots,e^{-j\pi(N_m^T-1)\sin\theta_m}\right]^{\top},\\
    \ba^{R}_{m}(\theta_m)=&\left[1,e^{-j\pi\sin\theta_m},\ldots,e^{-j\pi(N_m^R-1)\sin\theta_m}\right]^{\top}.
\end{align}
Then the vectorization $\bbf_m=\text{vec}(\bF_m)$ can be computed as
\begin{equation}
\label{eq:M radar vec}
\bbf_{m}=(\bI_{L}\otimes\bG_{mm})\bs_m+\sum_{m'\ne m}(\bI_{L}\otimes\bG_{mm'})\bs_{m'}+\bz_m,
\end{equation}
where $\bs_m = \text{vec}(\bS_m)$, and $\bz_{m}=\text{vec}(\bZ_{m})$. Letting $\hat\bbf_m=\bbf_m-(\bI_{L}\otimes\bG_{mm})\bs_m$, compute $\bK_m=\mathbb E[\hat\bbf_m\hat\bbf_m^{\text{H}}]$ as 
\begin{multline}
\bK_{m}  =\sum_{m'\ne m}\left(\bI_L \otimes \bG_{mm'} \right) \bs_{m'} \bs_{m'}^{\text{H}}\left(\bI_L \otimes \bG_{mm'} \right)^{\text{H}}\\+\sigma_{m}^2 \bI_{LN^R_m}.
\end{multline}
According to \cite{van2002optimum}, the \emph{Fisher information} about $\theta_m$ is
\begin{align}
J_m\left(\theta_m\right) & = 2\bs_m^{\text{H}}(\bI_L \otimes \dot{\bG}_{mm})^{\text{H}} \bK_{m}^{-1}(\bI_L \otimes \dot{\bG}_{mm}) \bs_m,
\end{align}
where $\dot{\bG}_{mm}=\partial \bG_{mm}/\partial \theta_m$. With each Cram\'{e}r-Rao bound computed as $1/J_m(\theta_m)$, we consider the following problem: 
\begin{subequations}
\label{prob:min sum of CRBs}
\begin{align}
\label{prob:min sum of CRBs:obj}
&\underset{\bs_m}{\text{minimize}}\quad\sum_{m=1}^{M}\frac{1}{J_m(\theta_m)}\\
&\text{subject to}\quad \|\bs_{m} \|^{2}\le P_{m}, \quad \forall m.
\end{align}
\end{subequations}

\subsubsection{Proposed FP-Based Method}

The above problem can be immediately written in the standard form in Theorem \ref{theorem:matrix_surrogate} as
\begin{subequations}
\label{prob:min sum of CRBs:trans}
\begin{align}
&\underset{\bs_m}{\text{maximize}}\quad\sum_{m=1}^{M}f^+_m\left(\sqrt{\bV}^{\text{H}}_{m} \bK_{m}^{-1}\sqrt{\bV}_m\right)\\
&\text{subject to}\quad \lVert \bs_{m} \rVert^{2}\leq P_{m}, \quad \forall m,
\end{align}
\end{subequations}
where $\sqrt{\bV}_m = \left(\bI_L \otimes \dot{\bG}_{mm}\right) \bs_m$ and $f_m^{+}(r) = -1/(2r)$. However, a subtle issue with the above matrix-FP problem is that the matrix denominator $\bK_m$ is not convex in $\bs_m$, so the optimization of $\bs_m$ is still difficult even after applying the unified quadratic transform. We propose addressing this issue by the Schur complement. Thus, problem \eqref{prob:min sum of CRBs:trans} is recast to
\begin{subequations}
\label{prob:min sum of CRBs:trans:equivalent}
\begin{align}
\underset{\bs_m,\,\bU_m}{\text{maximize}}&\quad\sum_{m=1}^{M}f^+_m\left(\sqrt{\bV}^{\text{H}}_{m} \bm\Lambda_{m}^{-1}\sqrt{\bV}_m\right)\\
\text{subject to}&\quad\; \|\bs_{m} \|^{2}\leq P_{m},\quad \forall m
    \label{prob:min sum of CRBs:trans:equivalent:cons1}\\
&\quad\; \bU_m\in\mathbb C^{LN^T_m \times LN^T_m},\quad\forall m\\
&\quad\left[\begin{array}{ll}
\bU_m & \bs_m \\
\bs_m^{\text{H}} & 1
\end{array}\right] \succeq \bm{0},\quad\forall m,
    \label{prob:min sum of CRBs:trans:equivalent:cons3}
\end{align}
\end{subequations}
where $\bm\Lambda_m = \sum_{m'\ne m}\left(\bI_L \otimes \bG_{m{m'}} \right) \bU_{m'}\left(\bI_L \otimes \bG_{mm'} \right)^{\text{H}}+\sigma_{m}^2 \bI$.
Applying the unified quadratic transform to the above problem yields the following reformulation:
\begin{subequations}
\label{prob:min sum of CRBs:equivalent}
\begin{align}
\underset{\bs_m,\,\bU_m,\,\bY_m}{\text{maximize}}&\quad\sum_{m=1}^{M}f^+_m\left(Q^+_m\right)\\
\text{subject to}&\quad\; \text{\eqref{prob:min sum of CRBs:trans:equivalent:cons1}--\eqref{prob:min sum of CRBs:trans:equivalent:cons3}}\\
&\quad\; \bY_m\in\mathbb C^{LN^T_m}, \quad\forall m,
\end{align}
\end{subequations}
where
\begin{equation}
    Q^+_m = \sqrt{\bV}_m^{H}\bY_m +\bY^{H}_m \sqrt{\bV}_m-\bY_m^{H}\bm\Lambda_m\bY_m.
    \label{}
\end{equation}
An efficient algorithm based on alternating optimization then readily follows. When the primal variables $\bs_m$ and $\bU_m$ are fixed, the auxiliary variable $\bY_m$ can be optimally updated as
\begin{equation}
    \bY_m=\bm\Lambda_m^{-1}\sqrt{\bV}_m.
    \label{eq:y_update:CRB}
\end{equation}
For fixed $\bY_m$, the new problem \eqref{prob:min sum of CRBs:equivalent} turns out to be jointly convex in $\bs_m$ and $\bU_m$ and thus can be efficiently solved by the standard method. This FP-based waveform design method is summarized in Algorithm \ref{algorithm:CRB}.

\begin{algorithm}[t]
  \caption{Waveform Design for Multi-Radar Sensing}
  \label{algorithm:CRB}
  \begin{algorithmic}[1]
      \STATE Initialize $\{\bs_m\}$ and $\{\bU_m\}$ to feasible values.
      \REPEAT 
      \STATE Update each $\bY_m$ by \eqref{eq:y_update:CRB}.
      \STATE Solve the convex problem of $(\bs_m,\bU_m)$ in \eqref{prob:min sum of CRBs:equivalent}.
      \UNTIL{the objective value converges} 
  \end{algorithmic}
\end{algorithm}
% where $Q^+_m = \sqrt{\bV}_m^{\text{H}}\bY_m +\bY^{\text{H}}_m \sqrt{\bV}_m-\bY_m^{\text{H}}\bm\Lambda_m\bY_m$ and $\{\bY_m\}$ is introduced by the unified quadratic transform. Following Algorithm \ref{algorithm:Iterative Algorithm for Matrix  Mixed FP}, we optimize $\{\bY_m\}$ and $\{\bs_m,\bU_m\}$ in an iterative manner. The optimal $\{\bY_m\}$ for fixed $\{\bs_m,\bU_m\}$ is 
% \begin{equation}
% \label{eq:y_update:CRB}
%     \bY^\star_m=\bm\Lambda_m^{-1}\sqrt{\bV}_m.
% \end{equation}
% Then, finding the optimal $\{\bs_m,\bU_m\}$ for fixed $\{\bY_m\}$ is a convex problem. This waveform design method for joint target sensing is summarized in Algorithm \ref{algorithm:CRB}. By Theorem \ref{theorem:convergence:UQT}, Algorithm \ref{algorithm:CRB} guarantees a convergence to a stationary point of the original problem \eqref{prob:min sum of CRBs}.

\subsubsection{Numerical Results}
\begin{figure}[t]
  \centering
  \includegraphics[width=8.5cm]{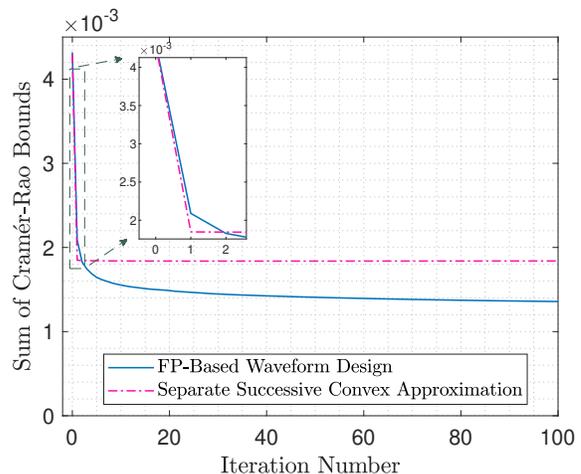}
  \caption{Minimizing the sum of the Cram\'{e}r-Rao bounds on the estimation error of DOA across $5$ radar sets when each $P_m=30$ dBm.}
  \label{fig:Min_sum_of_CRBs}
\end{figure}

\begin{figure}[t]
  \centering
  \includegraphics[width=8.5cm]{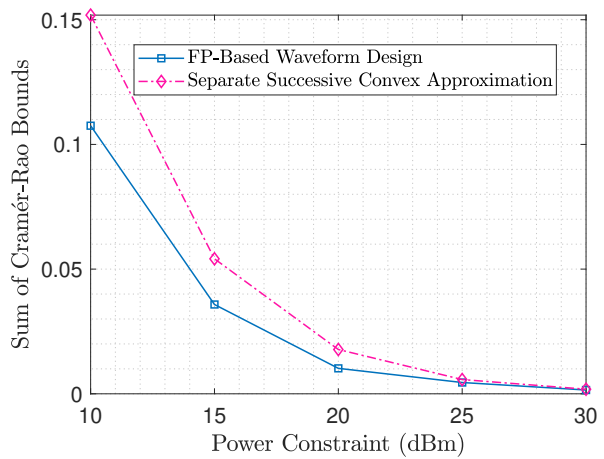}
  \caption{Sum of the Cram\'{e}r-Rao bounds on the estimation error of DOA across $5$ radar sets under the different power constraints.}
  \label{fig:different_p}
\end{figure}

The simulation setting follows. Consider $5$ radar sets with $L=4$. For $m=1,2,\ldots,5$, let $N^T_m$ in order be $(4,2,2,2,2)$, let $N^R_m$ in order be $(6,4,4,4,4)$, and let $\theta_m$ in order be $(\frac{1}{6}\pi,\frac{1}{3}\pi,\frac{1}{4}\pi,\frac{2}{5}\pi,\frac{3}{7}\pi)$. We let every $\beta_{mm'}=1$ and let every $P_m=P$, while trying different power constraints $P=10,15,\ldots,30$ dBm. We compare the proposed method with the existing algorithm in \cite{cheng2019co}. As shown in Fig.~\ref{fig:Min_sum_of_CRBs}, the benchmark algorithm converges faster than the proposed algorithm, and yet the latter ultimately achieves a much lower sum of the Cram\'{e}r-Rao bounds. the convergence of the proposed method is around $70\%$ lower than the starting point and is around $25\%$ lower than the convergence of the existing method in \cite{cheng2019co}. As shown in Fig.~\ref{fig:different_p}, the proposed FP-based method is superior to the benchmark under the other power constraints as well. In general, the stricter the power constraint, the greater advantage our algorithm has over the benchmark.

\subsection{Power Control for Secure Transmission}
\label{sec:secure transmission}

\subsubsection{Background} The aim of secure transmission is to send messages to legitimate receivers without any information leakage to eavesdroppers. Aside from the information-theoretical study, the optimization aspect has attracted extensive research interest as well. 
Considering a two-link interference channel with one eavesdropper, the authors of \cite{kalantari2015joint} devise a power control algorithm based on an altruistic and egoistic setting. For a wiretap channel with one legitimate receiver, one eavesdropper, and multiple jammers, \cite{wu2016secure} characterizes the power control as a Srtacjelberg game and thereby proposes an iterative optimization that achieves a Srtacjelberg equilibrium. Moreover, \cite{zhang2022secure} suggests a second-order-cone approach to power control for secure transmission in a cell-free MIMO network. For the UAV network case, \cite{zhang2019securing} proposes optimizing powers via block coordinate descent and successive convex approximation. %The use of machine learning has become popular in this area, e.g., the Q-learning approach in \cite{jameel2022secure}.

\subsubsection{Problem Formulation}
\begin{figure}[t]
  \centering
  \includegraphics[width=7cm]{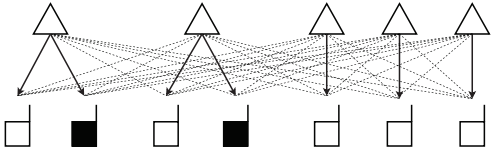}
  \caption{A multi-cell wiretap network with $L=5$ and $K=2$. The triangles are BSs, the white cellphones are legitimate users, the black cellphones are eavesdroppers, the solid arrows indicate cells, and the dashed lines represent cross-cell interference. Assume that each eavesdropper attacks its own cell.}
\label{fig:secrecy_rate_model}
\end{figure}

Consider $L$ base-stations (BSs) each serving a legitimate downlink user terminal. Without loss of generality, assume that the first $K$ ($K\le L$) BSs have one eavesdropper each. Fig.~\ref{fig:secrecy_rate_model} shows an example.  We use $p_i$ to denote the transmit power of the $i$th BS, $h_{ji}\in\mathbb C$ the channel from the $i$th BS to the $j$th legitimate user, $\tilde h_{ki}\in\mathbb C$ the channel from the $i$th BS to the $k$th eavesdropper, $\sigma^2_i$ the background noise power at the $i$th legitimate user, and $\tilde\sigma^2_k$ the background noise power at the $k$th eavesdropper. Assuming that cross-cell interference is treated as noise, the secret data rate of BS $i=1,\ldots,K$ with eavesdropper is given by
\begin{multline}
\label{secure_rate}
  R_i = \log\left(1+\frac{|h_{ii}|^2p_i}{\sum^L_{j=1,j\ne i}|h_{ij}|^2p_j+\sigma^2_i}\right)\\
  -\log\left(1+\frac{|\tilde h_{ii}|^2p_i}{\sum^L_{j=1,j\ne i}|\tilde h_{ij}|^2p_j+\tilde\sigma^2_i}\right),
\end{multline}
while the data rate of BS $i=K+1,\ldots,L$ without eavesdropper is given by
\begin{equation}
  R_i = \log\left(1+\frac{|h_{ii}|^2p_i}{\sum^L_{j=1,j\ne i}|h_{ij}|^2p_j+\sigma^2_i}\right).
\end{equation}
We seek the optimal power $\bp=(p_1,\ldots,p_L)$ under the power constraint $p_i\le P$
to maximize the sum weighted rates, i.e.,
\begin{subequations}
  \label{prob:secure transmission}
\begin{align}
\underset{\bp}{\text{maximize}} &\quad \sum^{L}_{i=1}w_iR_i
\label{obj:secure transmission}\\
\text{subject to}&\quad\; 0\le p_i\le P,\;\text{for}\; i=1,\ldots,L,
  \label{prob:secure transmission:cons_c}
\end{align}    
\end{subequations}
where $w_i$'s are the nonnegative weights. We remark that the above problem is a mixed max-and-min FP problem, so the existing quadratic transform in \cite{shen2018fractional} does not work for it.

\subsubsection{Direct FP-Based Method}

It is intriguing to apply the unified quadratic transform to the above problem because of the SINR terms. Unfortunately, those SINRs nested in the $-\log(\cdot)$ do not meet the requirement for $f_n^-$ as stated in Theorem \ref{theorem:UQT}. Nevertheless, this issue can be fixed by rewriting \eqref{secure_rate} as
\begin{multline}
\label{secure_rate:new}
  R_i = \log\left(1+\frac{|h_{ii}|^2p_i}{\sum^L_{j=1,j\ne i}|h_{ij}|^2p_j+\sigma^2_i}\right)\\
  +\log\left(1-\frac{|\tilde h_{ii}|^2p_i}{\sum^L_{j=1}|\tilde h_{ij}|^2p_j+\tilde\sigma^2_i}\right).
\end{multline}
We now treat each $\frac{|h_{ii}|^2p_i}{\sum^L_{j=1,j\ne i}|h_{ij}|^2p_j+\sigma^2_i}$ as the ratio nested in the concave increasing function $f^+_i(r)=\log(1+r)$, for $i=1,\ldots,L$, and treat each $\frac{|\tilde h_{kk}|^2p_k}{\sum^L_{j=1}|\tilde h_{kj}|^2p_j+\tilde\sigma^2_k}$ as the ratio nested in the concave decreasing function $f^-_k(r)=\log(1-r)$, for $k=1,\ldots,K$. Thus, the unified quadratic transform in Theorem \ref{theorem:UQT} immediately applies. The resulting new problem is
\begin{subequations}
  \label{prob:secure transmission:quadratic}
\begin{align}
\underset{p_i,y_i,\tilde y_k}{\text{maximize}} &\quad \sum^{L}_{i=1}\log(1+Q^+_i)+\sum^K_{k=1}\log\left(1-\frac{1}{Q^-_k}\right)
\label{prob:WTC:obj:new}\\
\text{subject to}&\quad\; 0\le p_i\le P,\;\text{for}\; i=1,\ldots,L\\
&\quad\; Q^-_k>0,\;\text{for}\; k=1,\ldots,K,
\end{align}    
\end{subequations}
where 
\begin{align*}
    Q^+_i &= 2y_i\sqrt{|h_{ii}|^2p_i}-y^2_i\left(\sum\nolimits_{j\ne i}|h_{ij}|^2p_j+\sigma^2_i\right),\\
     Q^-_k &= 2\tilde y_k\sqrt{\sum\nolimits_{j}|\tilde h_{kj}|^2p_j+\tilde\sigma^2_k}-\tilde y^2_k|\tilde h_{kk}|^2p_k.
\end{align*}
An iterative algorithm can be carried out for the above new problem. When $\bp$ is held fixed, the auxiliary variables are optimally updated as 
\begin{align}
  y_i &= \frac{\sqrt{|h_{ii}|^2p_i}}{\sum^L_{j=1,j\ne i}|h_{ij}|^2p_j+\sigma^2_i},
  \label{eq:secure transmission:y update}\\
  \tilde y_k &= \frac{\sqrt{\sum^L_{j=1}|\tilde h_{kj}|^2p_j+\tilde\sigma^2_k}}{|\tilde h_{kk}|^2p_k+\varepsilon}.
  \label{eq:secure transmission:yy update}
\end{align}
When the auxiliary variables are fixed, solving for $\bp$ in \eqref{prob:secure transmission:quadratic} is a convex problem. Algorithm \ref{algorithm:secure transmission:direct FP} summarizes this FP-based power control method.

\begin{algorithm}[t]
  \caption{Direct Power Control for Secure Transmission}
  \label{algorithm:secure transmission:direct FP}
  \begin{algorithmic}[1]
      \STATE Initialize $\bp$ to a feasible value.
      \REPEAT 
      \STATE Update each $y_i$ by \eqref{eq:secure transmission:y update} and update each  $\tilde{y}_i$ by \eqref{eq:secure transmission:yy update}.
      \STATE Solve the convex problem of $\bp$ in \eqref{prob:secure transmission:quadratic}.
      \UNTIL{the objective value converges}
  \end{algorithmic}
\end{algorithm}

\begin{algorithm}[t]
  \caption{Fast Power Control for Secure Transmission}
  \label{algorithm:secure transmission:LDT FP}
  \begin{algorithmic}[1]
      \STATE Initialize $\bp$ to a feasible value.
      \REPEAT 
      \STATE Update each  $\gamma_i$ by \eqref{eq:g_update:secure transimission} and update each  $\tilde{\gamma}_i$ by \eqref{eq:gg_update:secure transimission}.
      \STATE Update each  $y_i$ by \eqref{eq:y update:secure transmission:L_Q} and update each $\tilde{y}_i$ by \eqref{eq:yy update:secure transmission:L_Q}.
      \STATE Solve the convex problem of $\bp$ in \eqref{prob:secure transmission:LDT_FP}.
      \UNTIL{the objective value converges} 
  \end{algorithmic}
\end{algorithm}

\subsubsection{Fast FP-Based Method}

As formerly discussed in Section \ref{subsec:Lagrangian}, the generalized Lagrangian dual transform moves the ratios to the outside of logarithms and thereby improves the efficiency of numerical optimization. This strategy works for the present application case. By Theorem \ref{theorem:Lagrangian Dual Transform}, the original problem \eqref{prob:secure transmission} is converted to
\begin{subequations}
\label{prob:secure transmission:Lagrangian}
    \begin{align}
        \underset{\bp,\,\bm{\gamma},\,\tilde{\bm {\gamma}}}{\text{maximize}}&\quad f_r(\bp,\bm{\gamma},\tilde{\bm {\gamma}})
        \label{prob:secure transmission:Lagrangian:obj}\\
        \text{subject to}&\quad 0\le p_i\le P,\quad\forall i\\
        &\quad \gamma_i>0,\; \tilde{\gamma}_k>0,\;\forall i,k,
    \end{align}
\end{subequations}
where $\bm{\gamma}=\{\gamma_i\}^{L}_{i=1}$, $\tilde{\bm{\gamma}}=\{\tilde{\gamma}_i\}^{K}_{k=1}$, and the new objective function is
 \begin{align}
    &f_r(\bp,\bm {\gamma},\tilde{\bm {\gamma}})=\notag\\
    &\sum_{i=1}^{L}\left( w_i\log{(1+\gamma_i)}-w_i\gamma_i+\frac{w_i(1+\gamma_i)|h_{i,i}|^2p_i}{\sum_{j=1}^{L}|h_{i,j}|^2p_j+\sigma_i^2}\right)\notag\\
    &+\sum_{k=1}^{K}\left( w_k\log{(1-\tilde{\gamma}_k)}+w_k\tilde{\gamma}_k-\frac{w_k(1-\tilde{\gamma}_k)|\tilde{h}_{k,k}|^2p_k}{\sum_{j\ne k}|\tilde{h}_{k,j}|^2p_j+\tilde{\sigma}^2_k}\right).\notag
\end{align}
When $\bp$ is held fixed, $\bm\gamma$ and $\tilde{\bm\gamma}$ are optimally updated as 
\begin{align}
\gamma_i&=\frac{|h_{i,i}|^2p_i}{\sum_{j\ne i}|h_{i,j}|^2p_j+\sigma_i^2},
    \label{eq:g_update:secure transimission}\\
    \tilde{\gamma}_k &= \frac{|\tilde{h}_{k,k}|^2p_k}{\sum_{j=1}^{L}|\tilde{h}_{k,j}|^2p_j+\tilde{\sigma}^2_k}.
    \label{eq:gg_update:secure transimission}
\end{align}
We further consider optimizing $\bp$ in \eqref{prob:secure transmission:Lagrangian} for fixed $(\bm\gamma,\tilde{\bm\gamma})$; this task can be recognized as the sum-of-ratios problem. By the unified quadratic transform in Theorem \ref{theorem:UQT}, 
the problem of optimizing $\bp$ in \eqref{prob:secure transmission:Lagrangian} for fixed $(\bm\gamma,\tilde{\bm\gamma})$ is recast to 
\begin{subequations}
\label{prob:secure transmission:LDT_FP}
     \begin{align}  \underset{\bp,\,\by,\,\tilde{\by}}{\text{maximize}}&\quad \sum_{i=1}^{L}Q^+_i-\sum_{k=1}^{K}\frac{1}{Q_k^-}
        \label{obj:secure transmission:La_qua}\\
        \text{subject to}&\quad0\le p_i\le P,\quad\forall i\\
        &\quad Q^-_k>0,\quad \forall k\le K,
    \end{align}
\end{subequations}
where
\begin{align*} Q^+_i&=2y_i\sqrt{w_i(1+\gamma_i)|h_{i,i}|^2p_i}-y^2_i\left(\sum_{j=1}^{L}|h_{i,j}|^2p_j+\sigma^2_i\right),\\
    Q_k^-&={2\tilde{y}_k\sqrt{\sum_{j\ne k}|\tilde{h}_{k,j}|^2 p_j+\tilde{\sigma}^2_k}-\tilde{y}_k^2w_k(1-\tilde{\gamma}_k)|\tilde{h}_{k,k}|^2p_k}.
\end{align*}
When $\bp$ is fixed, the auxiliary variables $\by$ and $\tilde\by$ are respectively updated as
\begin{align}
    y_i =& \frac{\sqrt{w_i(1+\gamma_i)|h_{i,i}|^2p_i}}{\sum_{j=1}^{L}|h_{i,j}|^2p_j+\sigma^2_j},
    \label{eq:y update:secure transmission:L_Q}\\
    \tilde{y}_k =& \frac{\sqrt{\sum_{j\ne k}|\tilde{h}_{k,j}|^2 p_j+\tilde{\sigma}^2_k}}{w_k(1-\tilde{\gamma}_k)|\tilde{h}_{k,k}|^2p_k+\varepsilon}.
    \label{eq:yy update:secure transmission:L_Q}
\end{align}
Finally, with the above auxiliary variables $(\bm\gamma,\tilde{\bm\gamma},\by,\tilde\by)$ all held fixed, optimizing $\bp$ in \eqref{prob:secure transmission:LDT_FP} is a convex problem. 
% {without logarithms}. 
Algorithm \ref{algorithm:secure transmission:LDT FP} summarizes the above procedure. 

\subsubsection{Numerical Results}
% \begin{table}[htbp]
%   \centering
%   \caption{FP performance against optimial}
%     \begin{tabular}{lrrr}
%     \toprule
%           & \multicolumn{1}{c}{Fast FP} & \multicolumn{1}{c}{Direct FP} & \multicolumn{1}{c}{E-Search} \\
%     \midrule
%     Objective value & 4.90   & 4.90   & 5.69 \\
%     Running time (s) & 20.39 & 29.06 & 7524 \\
%     \bottomrule
%     \end{tabular}%
%   \label{tab:addlabel}%
% \end{table}%

\begin{figure}
  \centering
  \includegraphics[width=8.5cm]{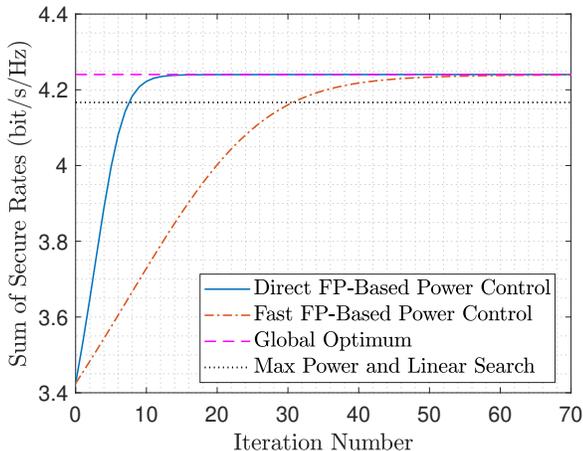}
  \caption{{Sum secure data rates vs. iteration number when $L=2$ and $K=2$.}}
  \label{fig:max_secure_rate_iteration}
\end{figure}

\begin{figure}[t]
  \centering
  \includegraphics[width=8.5cm]{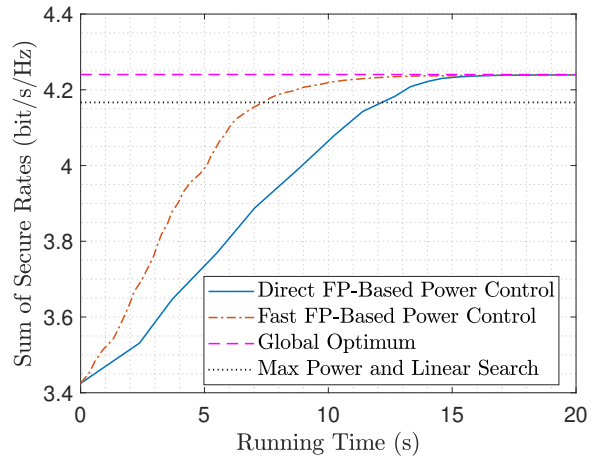}
  \caption{{Sum secure data rates vs. running time when $L=2$ and $K=2$; the simulation was done on a desktop computer with Intel i5-11500 processors.}}
  \label{fig:Max_sum_of_secure_rate}
\end{figure}

{We first evaluate the convergence. Let $L=2$, $K=2$, $\sigma_i^2=-10$ dBm, $\tilde\sigma_k^2=0$ dBm, and $P=10$ dBm. The channels are set as $|h_{11}|^2=1$, $|h_{12}|^2=0.1$, $|h_{21}|^2=0.09$, $|h_{22}|^2=0.87$, $|\tilde{h}_{11}|^2=0.5$, $|\tilde{h}_{12}|^2=0.11$,
$|\tilde{h}_{21}|^2=0.13$,
and $|\tilde{h}_{22}|^2=0.39$.} The power variables are all initialized to the max power at the starting point. We consider the following benchmark called \emph{max power and linear search}:  fix either $p_1$ or $p_2$ to the peak power and tune the other by linear search. We find the global optimum by an exhaustive search. {Fig.~\ref{fig:max_secure_rate_iteration} compares the convergence of the direct FP-based method and the fast FP-based method. The figure shows that the direct FP-based method and the fast FP-based method both converge to the global optimum. Notice that the direct FP converges faster; this result is expected because the direct FP approximates the original objective function only by Theorem \ref{theorem:UQT}, while the fast FP first approximates the original objective function by Theorem \ref{theorem:Lagrangian Dual Transform} and further approximates by Theorem \ref{theorem:UQT}. In other words, the approximation by the direct FP is tighter than that by the fast FP. Nevertheless, such a comparison is unfair because the per-iteration complexities of the two algorithms are quite different. Specifically, even though the direct FP and the fast FP both need to solve a convex problem iteratively, the former involves logarithms whereas the latter does not. For this reason, we believe that the running time is a more reasonable metric, especially when the same solver is used for the two methods.} Fig.~\ref{fig:Max_sum_of_secure_rate} compares the running time for the direct FP and the fast FP in solving the sum-of-secure-rates maximization problem. It can be seen that the fast FP has a much quicker convergence than the direct FP in terms of running time. {Thus, it is not true that the tighter approximation the better. Rather, a good approximation should be easy to solve for a numerical solver. Consequently, the fast FP is still more efficient even though its approximation is looser.}

\begin{figure}
  \centering \includegraphics[width=8.5cm]{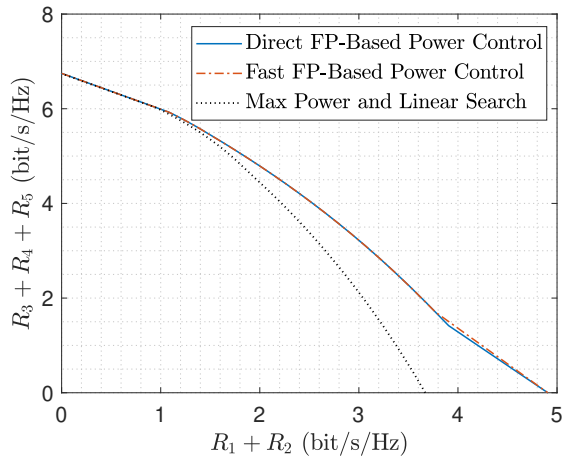}
\caption{{Tradeoff between the sum-rates $R_1+R_2$ at risk of eavesdropping and the sum-rates $R_3+R_4+R_5$ free of eavesdropping.}}
  \label{fig:tradeoff}
\end{figure}

We now show the tradeoff between the sum secure rates with eavesdropping and the sum rates without eavesdropping by trying out various rate weights $w_i$. Specifically, we fix $w_i$ to be 1 for $i\in\{1,2\}$, and set the rest $w_i$'s for $i\in\{3,4,5\}$ to the same value $\eta$ ranging from $0.001$ to $100$. The priority of secure transmission for the users at the risk of eavesdropping is inversely proportional to $\eta$. Let $L=5$, $K=2$, $\sigma_i^2=-10$ dBm, $\tilde\sigma_k^2=0$ dBm, and $P=10$ dBm. Let $|h_{ii}|^2=1,0.74,0.85,0.93,0.61$ for each $i=1,\ldots,5$; let $|\tilde{h}_{kk}|^2=0.50,0.15$ for each $k=1,2$; let $|h_{ji}|^2 = |\tilde{h}_{ki}|^2=0.1$ for each $i\ne j$ and $k\ne i$. We plot the tradeoff curve of the sum secure rates $R_1+R_2$ across the cells with eavesdropping versus the sum rates $R_3+R_4+R_5$ across the cells without eavesdropping. The Max power and linear search is extended as follows: let $p_1=p_2=\rho$ and let $p_3=p_4=p_5=\rho'$; fix either $\rho$ or $\rho'$ to the peak power and tune the other by linear search. It can be seen that the tradeoff curves of the two FP-based methods coincide. With the weights $w_i$ raised for $i\in\{3,4,5\}$, the sum rates without eavesdropping $R_3+R_4+R_5$ increases, while the sum rates with eavesdropping $R_1+R_2$ drops. The tradeoff curve of the benchmark method lies inside that of the proposed method, so its performance is inferior. For example, if $R_1+R_2$ is fixed at 3.4 bits/s/Hz, the FP-based method, either direct or fast, can enhance $R_3+R_4+R_5$ by around 150\% as compared to the benchmark method.

\section{Conclusion}
\label{sec:conclusion}

Differing from most of the previous studies in the literature, this work pursues a unified approach to both max-FP and min-FP. Specifically, we show that the existing quadratic transform for max-FP can be extended to min-FP, and then show that the new min-FP method and the existing max-FP method can be further integrated into a unified method that accounts for the mixed max-and-min FP problems. We then connect the proposed FP method to the MM theory, and in return {establish its convergence to a stationary point without requiring each subproblem to be uniquely solvable.} This MM interpretation also leads to a matrix extension of our method as well as a generalized Lagrangian dual transform that facilitates the log-ratio problem solving. Moreover, the paper illustrates the use of this new FP technique in minimizing the AoIs, in minimizing the Cram\'{e}r-Rao bounds on sensing error, and in maximizing the secure data rates. Many other possible applications can be envisioned, such as latency minimization, exposure does management of radiation, and ISAC.

% \section*{Acknowledgment}

% The preferred spelling of the word ``acknowledgment'' in America is without 
% an ``e'' after the ``g''. Avoid the stilted expression ``one of us (R. B. 
% G.) thanks $\ldots$''. Instead, try ``R. B. G. thanks$\ldots$''. Put sponsor 
% acknowledgments in the unnumbered footnote on the first page.

% \bibColoredItems{blue}{bertsekas1999nonlinear,zappone2017globally,matthiesen2020globally}

\appendix

\section*{Proof of Theorem \ref{theorem:matrix_surrogate}}
% {With a slight abuse of terminology}, 
We use $f_o(\bx)$ to denote the objective function in \eqref{prob:FP_mixed_matrix}. Our main goal is to show that $g(\bx|\hat\bx)$ in \eqref{matrix_g} is a surrogate function of $f_o(\bx)$. The rest results in Theorem \ref{theorem:matrix_surrogate} then immediately follow by the MM theory.

For ease of notation, define two new variables
\begin{align}
\bm{\Delta}_n&=\bY_n(\hat\bx)-\bB_n^{-1}(\bx)\sqrt{\bA_n}(\bx),\\
\tilde{\bm\Delta}_n &= \tilde\bY_n(\hat\bx)-\bA_n^{-1}(\bx)\sqrt{\bB_n}(\bx),
\end{align}
with which $g(\bx|\hat\bx)$ in \eqref{matrix_g} can be rewritten as
% \begin{align}
% &g(\bx|\hat\bx)\notag\\
% &=\sum_{n\le N_0}f_n^{+}\big(\sqrt{\bA}^{\text{H}}_n(\bx)\bB^{-1}_n(\bx)\sqrt{\bA}_n(\bx)-\bm{\Delta}^{\text{H}}_n\bB_n(\bx)\bm{\Delta}_n\big)\,+\notag\\
% &\,\sum_{n>N_0}f_n^{-}\big({\big(\text{max}\{\sqrt{\bB}^{\text{H}}_n(\bx)\bA^{-1}_n(\bx)\sqrt{\bB}_n(\bx)-\tilde{\bm\Delta}^{\text{H}}_n\bA_n(\bx)\tilde{\bm\Delta}_n,\bm 0^+\}\big)^{-1}}\big).\notag
% \end{align}
\begin{align}
&g(\bx|\hat\bx)=\notag\\
&\sum_{n\le N_0}f_n^{+}\big(\sqrt{\bA}^{\text{H}}_n(\bx)\bB^{-1}_n(\bx)\sqrt{\bA}_n(\bx)-\bm{\Delta}^{\text{H}}_n\bB_n(\bx)\bm{\Delta}_n\big)\,+\notag\\
&\sum_{n>N_0}f_n^{-}\big(\big({\sqrt{\bB}^{\text{H}}_n(\bx)\bA^{-1}_n(\bx)\sqrt{\bB}_n(\bx)-\tilde{\bm\Delta}^{\text{H}}_n\bA_n(\bx)\tilde{\bm\Delta}_n}\big)^{-1}\big).\notag
\end{align}
Because each $f^+_n(\cdot)$ is monotonically increasing with the input matrix, the first sum component on the right-hand side of the above equation can be bounded from above as
\begin{align}
     \sum_{n\le N_0}f_n^{+}\big(\sqrt{\bA}^{\text{H}}_n(\bx)\bB^{-1}_n(\bx)\sqrt{\bA}_n(\bx)-\bm{\Delta}^{\text{H}}_n\bB_n(\bx)\bm{\Delta}_n\big)\notag\\
{\le}\sum_{n\le N_0}f_n^+\left(\sqrt{\bA}^{\text{H}}_n(\bx)\bB^{-1}_n(\bx)\sqrt{\bA}_n(\bx)\right).
\label{bound:sum1}
\end{align}
Moreover, the second sum component is bounded as
% \begin{align}
% &\sum_{n>N_0}f_n^{-}\big({\big(\text{max}\{\sqrt{\bB}^{\text{H}}_n(\bx)\bA^{-1}_n(\bx)\sqrt{\bB}_n(\bx)-\tilde{\bm\Delta}^{\text{H}}_n\bA_n(\bx)\tilde{\bm\Delta}_n,\bm 0^+\}\big)^{-1}}\big)\notag\\
% &\overset{(a)}{\le}\sum_{n>N_0}f_n^-\big({\big(\text{max}\{\sqrt{\bB}^{\text{H}}_n(\bx)\bA^{-1}_n(\bx)\sqrt{\bB}_n(\bx),\bm 0^+\}\big)^{-1}}\big)\notag\\
% &\overset{(b)}{=}\sum_{n>N_0}f_n^-\big(\sqrt{\bA}^{\text{H}}_n(\bx)\bB^{-1}_n(\bx)\sqrt{\bA}_n(\bx)\big),
% \label{bound:sum2}
% \end{align}
\begin{align}
&\sum_{n>N_0}f_n^{-}\big(\big({\sqrt{\bB}^{\text{H}}_n(\bx)\bA^{-1}_n(\bx)\sqrt{\bB}_n(\bx)-\tilde{\bm\Delta}^{\text{H}}_n\bA_n(\bx)\tilde{\bm\Delta}_n}\big)^{-1}\big)\notag\\
&\overset{(a)}{\le}\sum_{n>N_0}f_n^-\big(\big({\sqrt{\bB}^{\text{H}}_n(\bx)\bA^{-1}_n(\bx)\sqrt{\bB}_n(\bx)\big)^{-1}}\big)\notag\\
&\overset{(b)}{=}\sum_{n>N_0}f_n^-\big(\sqrt{\bA}^{\text{H}}_n(\bx)\bB^{-1}_n(\bx)\sqrt{\bA}_n(\bx)\big),
\label{bound:sum2}
\end{align}
where $(a)$ follows by the decreasing monotonicity of $f^-_n(\cdot)$ and $(b)$ follows by the property assumed in \eqref{property:f^-}. Combining \eqref{bound:sum1} and \eqref{bound:sum2} gives $g(\bx|\hat\bx)\le f_o(\bx)$.

Furthermore, notice that the equality in $(a)$ holds if and only if every $\tilde{\bm\Delta}_n=0$ so that $\bx=\hat\bx$. Similarly, the equality in \eqref{bound:sum1} holds if and only if $\bx=\hat\bx$. To summarize, $g(\hat\bx|\hat\bx)=f_o(\hat\bx)$. The proof is then completed.

\bibliographystyle{IEEEtran}
\bibliography{IEEEabrv,refs}

\label{sec:refs}

\begin{IEEEbiographynophoto}{Yannan Chen}(Student Member, IEEE) received the B.E. degree in Automation Engineering and the M.E. degree in Pattern Recognition and Intelligent Systems from Xiamen University, Xiamen, China, in 2018 and 2021, respectively. He is currently pursuing his Ph.D. degree with the School of Science and Engineering at The Chinese University of Hong Kong (Shenzhen). His research interests include optimization, wireless communications, and machine learning.
\end{IEEEbiographynophoto}

\begin{IEEEbiographynophoto}{Licheng Zhao} received the B.S. degree in Information Engineering from Southeast University (SEU), Nanjing, China, in 2014, and the Ph.D. degree with the Department of Electronic and Computer Engineering at the Hong Kong University of Science and Technology (HKUST), in 2018. Since June 2018, he has been an algorithm engineer in recommendation system with JD.COM, China. Since Dec. 2021, he has served as a research scientist in Shenzhen Research Institute of Big Data (SRIBD).  His research interests are in optimization theory and efficient algorithms, with applications in signal processing, machine learning, and deep learning in recommendation system.
\end{IEEEbiographynophoto}

\begin{IEEEbiographynophoto}{Kaiming Shen}(Member, IEEE) received the B.Eng. degree in information security and the B.Sc. degree in mathematics from Shanghai Jiao Tong University, China in 2011, and then the M.A.Sc. degree in electrical and computer engineering from the University of Toronto, Canada in 2013. After working at a tech startup in Ottawa for one year, he returned to the University of Toronto and received the Ph.D. degree in electrical and computer engineering in early 2020. Since 2020, Dr. Shen has been with the School of Science and Engineering at The Chinese University of Hong Kong (Shenzhen), China as a tenure-track assistant professor. His research interests include optimization, wireless communications, information theory, and machine learning.

Dr. Shen received the IEEE Signal Processing Society Young Author Best Paper Award in 2021 for his work on fractional programming for communication systems. Dr. Shen serves as an Editor of IEEE Transactions on Wireless Communications.
\end{IEEEbiographynophoto}

\end{document}